\documentclass[twocolumn]{aastex62}
\usepackage{apjfonts} 
\received{2018 February 22 }
\revised{2018 June 12 }
\accepted{2018 June 12}
\shorttitle{The AEGIS Survey: The Evolution of Carbon and Iron}
\shortauthors{Yoon, J. et al.}

\begin{document}

\title{Galactic Archeology with the AEGIS Survey: The Evolution of Carbon and Iron in the Galactic Halo}
\author[0000-0002-4168-239X]{Jinmi Yoon} 
\author[0000-0003-4573-6233]{Timothy C. Beers}
\author{Sarah Dietz}
\affiliation{Department of Physics, University of Notre Dame, Notre Dame, IN 46556, USA; jinmi.yoon@nd.edu}
\affiliation{Joint Institute for Nuclear Astrophysics -- Center for the Evolution of the Elements (JINA-CEE), USA}
\author{Young Sun Lee}
\affiliation{Department of Astronomy and Space Science, Chungnam National University, Daejeon 34134, Korea}
\author[0000-0003-4479-1265]{Vinicius M. Placco}
\affiliation{Department of Physics, University of Notre Dame, Notre Dame, IN 46556, USA; jinmi.yoon@nd.edu}
\affiliation{Joint Institute for Nuclear Astrophysics -- Center for the Evolution of the Elements (JINA-CEE), USA}
\author{Gary Da Costa}
\author{Stefan Keller} 
\author{Christopher I. Owen}
\affiliation{Research School of Astronomy and Astrophysics,
Australian National University, Canberra, ACT 0200, Australia }
\author[0000-0001-9927-5255]{Mahavir Sharma}
\affiliation{Department of Physics, Institute for Computational Cosmology, Durham University, South Road, Durham DH1 3LE, UK}

\begin{abstract}
Understanding the evolution of carbon and iron in the Milky Way's halo is of importance because these two elements play crucial roles constraining star formation, Galactic assembly, and chemical evolution in the early Universe. Here, we explore the spatial distributions of carbonicity, [C/Fe], and
metallicity, [Fe/H], of the halo system based on
medium-resolution ($R \sim$ 1,300) spectroscopy of $\sim$58,000 stars in the Southern
Hemisphere from the AAOmega Evolution of Galactic Structure (AEGIS)
survey. The AEGIS carbonicity map exhibits a positive gradient with
distance, as similarly found for the Sloan Digital Sky Survey (SDSS) carbonicity map of Lee et al.  The metallicity map confirms that [Fe/H] decreases with distance,
from the inner halo to the outer halo. 
We also explore the formation and chemical-evolution history of the halo by considering the populations of carbon-enhanced
metal-poor (CEMP) stars present in the AEGIS sample. 
The cumulative and differential frequencies of CEMP-no stars (as classified by their characteristically lower levels of absolute carbon abundance, $A$(C) $\leq$ 7.1 for sub-giants and giants) increases with decreasing metallicity, and is \textit{substantially higher than previous  determinations} for CEMP stars as a whole. In contrast, that of CEMP-$s$ stars (with higher $A$(C)), remains almost flat, at a value $\sim$10\%, in the range $-\,4.0 \lesssim$ [Fe/H] $\lesssim-$2.0.  The distinctly different behaviors of the CEMP-no and CEMP-$s$ stars relieve the tension with 
population-synthesis models assuming a binary mass-transfer origin, which previously struggled to account for the higher
reported frequencies of CEMP stars, taken as a whole, at low metallicity. 

\end{abstract}


\section{Introduction}\label{introduction}

The present chemical composition of stars in the Milky Way (with the
exception of hydrogen and helium) comprises various nucleosynthetic
products, forged in previous generations of stars. First-generation
stars, which are expected to be massive stars formed from primordial
gas, synthesized metals up to the iron peak via stages of stellar
nucleosynthesis in their interiors, and (possibly) elements such as Sr or Ba via a weak slow neutron-capture process \citep[weak $s$-process, e.g.,][]{maeder2015,frischknecht2016}. 
Iron-peak elements are also created via explosive nucleosynthesis 
\citep[e.g.,][and references therein]{nomoto2013} by supernovae associated with massive stars.
Beyond the iron peak, roughly half of the heavy elements are
produced in low- to intermediate-mass stars by the main $s$-process
during the asymptotic giant-branch (AGB) phase 
\citep[e.g.,][]{frost1996,lugaro2003,herwig2005,karakas2014}. 
The so-called ``intermediate" neutron-capture process or $i$-process 
(possibly operating in high-mass AGB stars) may also play a role 
\citep{cowan1977, dardelet2015, hampel2016}. Other heavy
metals beyond the Fe-peak are created by the main rapid neutron-capture process ($r$-process), likely associated with
neutron star mergers \citep[e.g.,][and references therein]{lattimer1974,
meyer1989, rosswog2014, abbott2017a, drout2017,shappee2017}, but could also 
involve so-called magneto-rotational instability or "jet" supernovae 
\citep{cameron2003, fujimoto2008, winteler2012}, or neutrino-driven winds in core-collapse supernovae
\citep[][and references therein]{arcones2013}. 
In addition, there is another process, referred to as the weak or limited $r$-process \citep{travaglio2004,wanajo2006, frebel2018},
whose astrophysical site(s) are not yet clear, but it is thought to be associated with supernovae origins \citep{izutani2009,nomoto2013}. This process can explain the moderate enhancements of light neutron-capture elements such as  Sr, Y, and Zr relative to elements heavier than Ba, a signature that appears distinct from other neutron-capture processes (see, e.g., \citealt{honda2007}). 

All of the elements play potentially important roles in our
understanding of Galactic chemical evolution (GCE), since the production history
of each element can follow different nucleosynthesis pathways (exploring different
astrophysical processes, sites, timescales, and/or stellar-progenitor masses). However, in this 
work we focus on two fundamental elements, carbon and iron. These two
elements are of special significance, because they serve as tracers of the
stellar populations that were present from the earliest times in the
chemical evolution of the Galaxy.

\subsection{Carbon as a Tracer of Stellar Populations and GCE}

The observed abundances of most of the light and heavy elements in stars scale with the overall metallicity. However, as pointed out by \citet{beers1992}, carbon (and a number of other light elements, including N and O) is a notable exception. An increasing fraction of low-metallicity stars exhibit carbon enhancement with declining metallicity, approaching 100\% at the lowest iron abundances \citep{placco2014c}.

In the very early universe (likely within the first few hundred million years following the Big Bang), carbon is thought to be ejected primarily by so-called ``faint" supernovae \citep[e.g.,][]{umeda2003, umeda2005, nomoto2013, tominaga2014} of massive first-generation stars, by the stellar winds from massive, rapidly rotating spinstars \citep[e.g.,][]{meynet2006,meynet2010, chiappini2013}, and  by core-collapse supernovae from massive stars. Pollution
of the surrounding pristine interstellar medium (inside and outside the natal clouds of the first stars) by carbon provided pathways for efficient gas cooling and fragmentation, enabling the formation of low- and intermediate-mass stars \citep[e.g.,][]{bromm2003,schneider2003,schneider2012,omukai2005,frebel2007b}.

The progeny of the very first stars are expected to exhibit extremely low iron (and other heavy-element) content and greatly enhanced carbon. This first-star nucleosynthetic signature is matched by the sub-class of carbon-enhanced metal-poor (CEMP;\footnote{There are several CEMP ([Fe/H]$< -$1.0, [C/Fe] $\geq$ +0.7) sub-classes depending on enhancement of heavy neutron-capture elements.  
\newline CEMP-$s$ : [C/Fe] $\geq$ +0.7, [Ba/Fe] $>$ +1.0, and
[Ba/Eu] $>$ +0.5\newline CEMP-$r$ : [C/Fe] $\geq$ +0.7 and [Eu/Fe]$>$ +1.0
\newline CEMP-$i\,(r/s)$  : [C/Fe] $\geq$ +0.7 and 0.0 $<$ [Ba/Eu] $<$+0.5 
\newline CEMP-no : [C/Fe] $\geq$ +0.7 and [Ba/Fe] $<$ 0.0} 
\citealt{beers2005}; \citealt{aoki2007}) stars known as CEMP-no stars \citep[e.g.,][and references therein]{christlieb2004, meynet2006, frebel2008, nomoto2013,keller2014, bonifacio2015,yoon2016,placco2016b,chiaki2017,choplin2017}. 

Beginning roughly a Gyr later, the dominant carbon-production pathway is replaced by AGB nucleosynthesis in intermediate- and lower-mass stars.  This nucleosynthetic signature (an enhancement of both carbon and $s$-process elements) can be preserved on the surfaces of long-lived low-mass binary companions following a mass-transfer event from the erstwhile AGB stars \citep[e.g.,][]{lugaro2012,placco2013}. The CEMP-$s$ (and possibly CEMP-$i$, \citealt{hampel2016}) stars found at extremely and very low metallicity (but so far not at the lowest metallicity, [Fe/H]$< -$4.0) are the living records of this era.  

Nature's dual carbon-production pathways in cosmic time were first recognized as high and low bands of absolute carbon abundance, $A$(C)\footnote{$A$(C) = $\log\,\epsilon$(C) = $\log\,$($N_{\rm C}/N_{\rm H}$)+12, where $N_{\rm C}$ and $N_{\rm H}$ represent number-density fractions of carbon and hydrogen, respectively.}, in the $A$(C) vs. [Fe/H] space \citep{spite2013}, based on a sample of $\sim 50$ ``un-mixed" turnoff stars.  This behavior was supported  by \citet{bonifacio2015}, based on $\sim$70 CEMP stars, including a number of mildly evolved sub-giants. The full richness of the behavior of CEMP stars in this space was revealed in the Figure 1 of \citet{yoon2016} -- the Yoon-Beers diagram -- based on a large  literature sample of $\sim$300  CEMP stars with available high-resolution spectroscopy. Not only did this diagram identify two primary peaks in the 
marginal plot of $A$(C) (at $A$(C) $\sim$ 6.3 and 7.9), but Yoon et al. were able to sub-classify the CEMP stars into three primary groups, based on the morphology of CEMP stars in the $A$(C)-[Fe/H] diagram. In particular, the stars formerly referred to as ``carbon normal" by Spite et al. and Bonifacio et al. were shown to be CEMP stars that did not follow the ``band structure" as originally recognized. Instead, 
Yoon et al. identified the great majority of CEMP-$s$ stars as members of CEMP Group I stars, based on their distinctively higher $A$(C) compared to the CEMP-no stars, while most CEMP-no stars were classified as either CEMP Group II or Group III stars.  The Group II stars exhibited a strong dependency of $A$(C) on [Fe/H], while the Group III stars showed no such dependency. These different behaviors were also reflected by clear differences between Group II and Group III stars in the $A$(Na)-$A$(C) and $A$(Mg)-$A$(C) spaces (Figure 4 of Yoon et al.). Some of these apparent differences also appeared in recent theoretical work. For instance, \citet{sarmento2017} explored the Pop III enrichment of CEMP-no stars using the RAMSES cosmological simulation. One of their predictions clearly shows the presence of patterns visible in [C/H]-[Fe/H] space that might be associated with the Group II and III stars (their Fig 13). At the time, they were not aware of these groups and did not have a full sample of CEMP-no stars to compare with, thus they did not call attention to this result. However, they now agree that two different groups of CEMP-no stars indeed exist in their simulation predictions (R. Sarmento and E. Scannapieco, priv. comm.). A GADGET cosmological simulation by \citet{jeon2017}, which studied the chemical signature of Pop III stars, shows that there are indeed two groups of CEMP-no stars (M. Jeon, priv. comm.). 

The distinctively different patterns among the CEMP-no stars noted by Yoon et al. provided a first indication of possible multiple progenitors and/or the environments in which they formed. This has led to further exploration of the impact of dust cooling by grains of different compositions, e.g., carbon- vs. silicate-based dust \citep{chiaki2017}, to account for the formation of the Group II and III CEMP-no stars. 

Finally, Yoon et al. demonstrated that the carbon bi-modality in the marginal plot of absolute carbon abundance histogram in the Yoon-Beers diagram could be used to separate the CEMP-no stars from CEMP-$s$ stars based on $A$(C) alone, which can be readily obtained from medium-resolution spectra (dividing at $A$(C) = 7.1), at a success rate similar to that obtained using the [Ba/Fe] ratios (as defined by \citealt{beers2005}), which generally require high-resolution spectroscopy to measure. This opens the possibility to explore the global properties of the populations of CEMP-no and CEMP-$s$ stars from the already very numerous medium-resolution spectra that are available for CEMP stars, as we do in this work.

\subsection{Iron as a Tracer of Stellar Populations and GCE}

In the early universe, iron was synthesized mainly by core-collapse supernovae from massive stars. Later ($\sim 1$ Gyr after the Big Bang), the dominant production pathway of iron changed to Type Ia supernovae, associated with thermonuclear explosions of C+O white dwarfs \citep[e.g.,][]{mcwilliam1997,frebel2013}. The abundance of iron is often taken to represent the overall metallicity in stars, since iron has the highest number density among the heavy metals and is predominantly observed in metal-poor stars. Thus, the iron-to-hydrogen ratio ([Fe/H]; often interchangeably used with metallicity) is another crucial probe of stellar populations. The spatial metallicity distribution function (MDF) provides a record of the metal-enrichment history for different populations and in different regions of the Galaxy. In addition, [Fe/H] can also serve as a rough, indirect age proxy (except in the lowest-metallicity regime, where local inhomogeneous enrichment dominates \citep[e.g.,][]{kobayashi2011,el-badry2018}. 

\subsection{Outline of this Paper}

Previous work has been based on small samples of halo stars with
available high-resolution spectroscopic abundance determinations 
\citep[e.g.,][]{barklem2005, aoki2013, norris2013b, roederer2014}, or much larger samples of stars with medium-resolution spectroscopy, primarily in the Northern Hemisphere (e.g., the Sloan Digital Sky Survey; SDSS \citealt{york2000}; \citealt{yanny2009}; and the Large Sky Area Multi-Object Fiber Spectroscopic Telescope survey; LAMOST, \citealt{cui2012}). 

In this paper we make use of a new large sample of stars in the Southern
Hemisphere to consider several important probes of the chemical
evolution and assembly history of the Galaxy. Section \ref{sec:data}
briefly describes the medium-resolution spectroscopic data obtained by
the AAOmega Evolution of Galactic Structure (AEGIS) survey (P.I. Keller). We then present our results on
the spatial distributions of [C/Fe] (carbonicity) 
and metallicity in Section~\ref{sec:cartography}. Section \ref{sec:cemp} 
considers the CEMP stars in the AEGIS survey, separated into CEMP-no and CEMP-$s$ stars
based on $A$(C). In
Section~\ref{sec:frequency} we explore the cumulative and differential
frequencies of the CEMP stars. Section~\ref{sec:discussion} describes 
implications for the chemical evolution and formation history of 
the Galactic halo system, based on the results reported in 
Section~\ref{sec:cartography} --~\ref{sec:frequency}. We conclude with a summary and description of future work in Section~\ref{sec:future}. 

\section{Data}\label{sec:data}

While the SDSS, in particular its
stellar-specific programs, the Sloan Extension for Galactic Understanding and Exploration \citep[SEGUE-1 and SEGUE-2;][]{yanny2009}, 
has greatly advanced our understanding of the chemical evolution and assembly
history of the Galaxy, no
extensive wide-angle spectroscopic surveys in the Southern Hemisphere
existed prior to AEGIS. Although the HK Survey of \citet{beers1985} and
\citet{beers1992} and the Hamburg/ESO Survey of Christlieb and
colleagues \citep{christlieb2003} obtained medium-resolution
spectroscopic follow-up for some 20,000 candidate metal-poor stars, these were very sparsely distributed over the southern sky, and left large swaths of sky completely unsampled. We briefly introduce the AEGIS program below.\\

\begin{figure*}[thb!]
\centering
\includegraphics[scale=0.8]{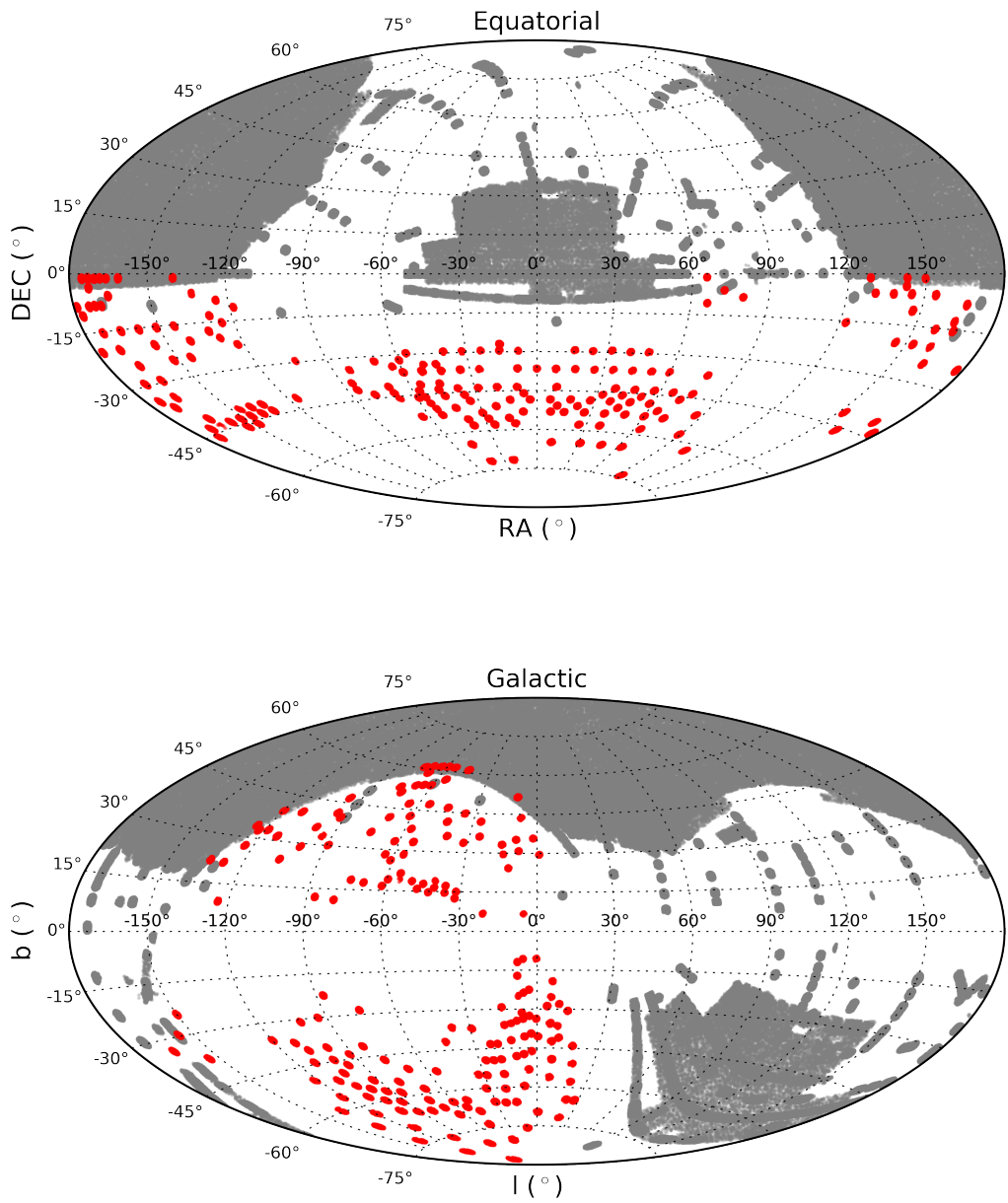}
\caption{Aitoff projections of the SDSS footprint (gray)
and the AEGIS footprint (red) in the equatorial coordinate system (upper
panel) and in the Galactic coordinate system (lower panel). 
\label{ffootprint}}
%\vspace{-11.6064pt}
\end{figure*}

The AEGIS survey is a 
medium-resolution ($R \sim$1,300) spectroscopic survey in the Southern
Hemisphere, with the goal to determine the chemistry and kinematics of
thick-disk and halo stars in order to constrain the chemodynamical
evolution of the Milky Way. The input catalog for the spectroscopic targets was derived from photometric observations of a set of approximately 2-degree diameter fields taken during commissioning of the SkyMapper telescope \citep{keller2007}.  The gravity 
and metallicity sensitivity of the SkyMapper photometric system \citep{keller2007} allowed the focus of the target catalog to be on blue horizontal-branch, red clump, and metal-poor star candidates.  The AEGIS sample excludes the region of sky within a 10 deg radius of the Galactic Center, and, in addition, a small number of candidate extremely metal-poor stars that formed the basis for a separate follow-up program \citep[e.g.,][]{jacobson2015}.  

Spectroscopic observations were carried out using the AAOmega multi-fibre dual-beam spectrograph \citep[e.g.,][]{sharp2006} on the 3.9m Anglo-Australian Telescope (AAT).  Spectra were obtained for a total of $\sim$70,000 stars distributed over 4,900 square degrees of the southern sky during the four semesters of allocated time.  All the survey observations were run through a uniform data reduction process based on the 2DFDR reduction code\footnote{\url{https://www.aao.gov.au/science/software/2dfdr}}.  Here we make use of the blue-arm spectra which, with the 580V grating, yields a wavelength coverage of approximately $\lambda$3750-5400 \AA\, and a resolving power $R \approx$ 1,300.  A more complete description of the AEGIS data, sample spectra, and the analysis techniques used to derive the atmospheric parameters, as well as estimates of the [C/Fe] ratios and (photometric) distances, is provided in the Appendix\footnote{Note that, even though the Appendix describes the corrections we apply to the measured atmospheric parameters, for simplicity in the remainder of this paper, we employ the notation for effective temperature, $T_{\rm eff}$, surface gravity, log $g$,
metallicity, [Fe/H]), and carbonicity, [C/Fe]; the corrections have been made for all of these parameters, as appropriate.}.

Figure~\ref{ffootprint} compares the footprints on the sky of stars observed during SDSS/SEGUE and AEGIS.  The sky coverage for SDSS/SEGUE was nearly contiguous over large portions of the sky, due to the
numerous calibration stars (photometrically selected to be likely metal-poor main-sequence turnoff stars) observed during the extragalactic programs carried out during operation of the SDSS. The AEGIS footprint is rather sparse, but it covers the regions
that SDSS could not reach.

\begin{figure*}[th!]
\centering
\gridline{\fig{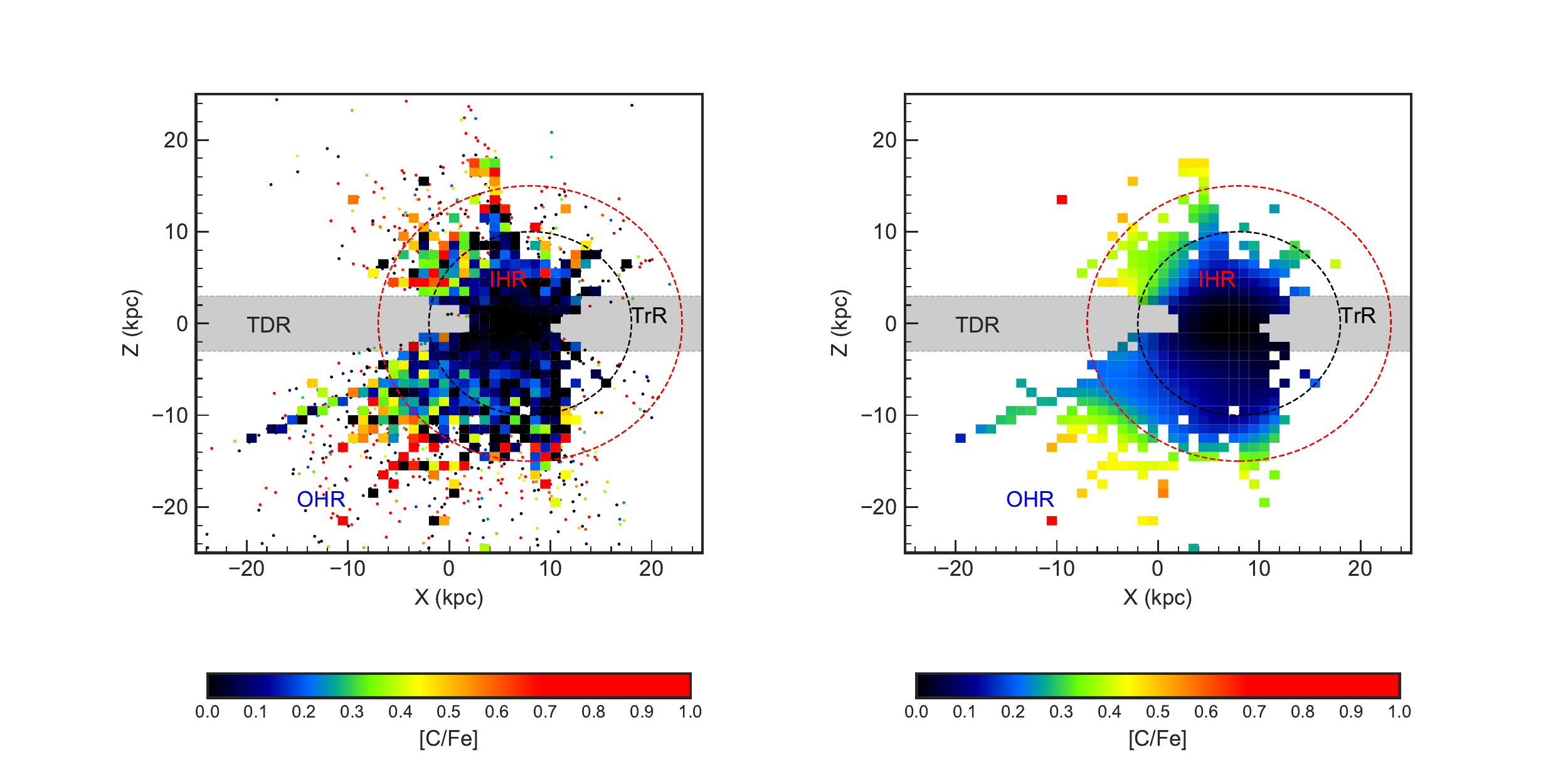}{0.9\textwidth}{(a) Carbonicity map}}
\gridline{\fig{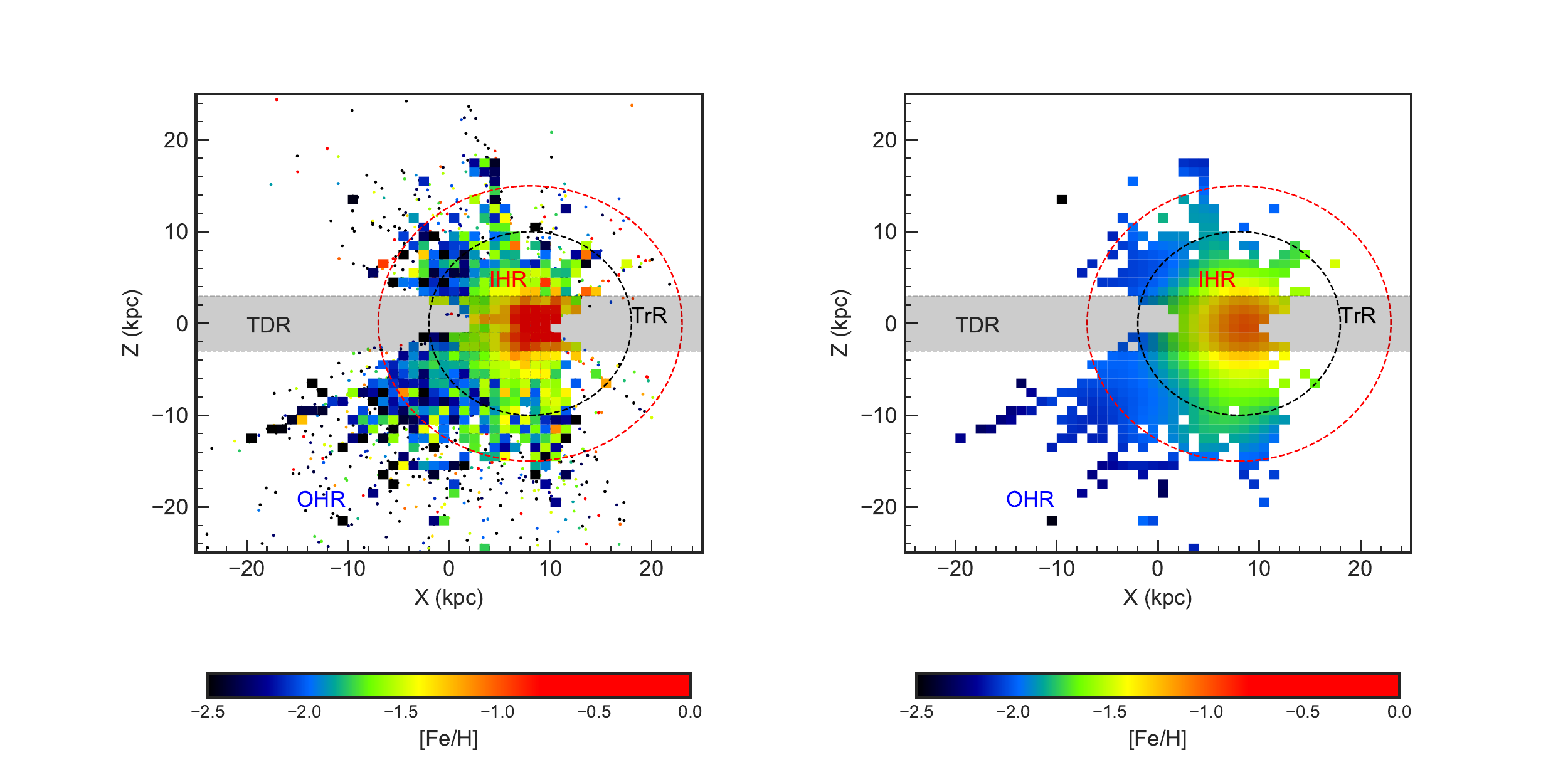}{0.9\textwidth}{(b) Metallicity map}}
\caption{Cartographic maps of carbonicity (upper panels) and metallicity
(lower panels) projected onto the X-Z plane. 
The adopted Cartesian reference system is right-handed, having
positive X towards the anti-center of the Galaxy. The location at (0,0) kpc corresponds to the center of the Galaxy, and the Sun is located at
(X, Z) = (8.0, 0.0) kpc. The left panels
show the distribution of stars in the X-Z plane in a square grid  of 0.5 kpc $\times$ 0.5 kpc pixels. The filled squares have pixels with at least three stars and
the filled dots indicate pixels with two or one stars. The right panels
show the stellar distributions in the pixel grid
smoothed with a two-pixel Gaussian kernel. The colors in the upper
panels and the bottom panels represent the median [C/Fe] and [Fe/H]
values at each pixel, respectively, as shown by the color bar under each panel.  The  upper and lower limits shown in the color bar are also used for pixels whose [C/Fe] or [Fe/H] lie above or below these limits. The gray-shaded area
indicates the thick-disk region (TDR). The black- and red-dashed circles
represent the dividing lines for the inner-halo (IHR), the transition
region (TrR), and the outer-halo (OHR), based on the distribution of
carbonicity. See text for the definition of these regions.
\label{fmapsxz}}
\end{figure*}

\begin{figure*}%[th!]
\gridline{\fig{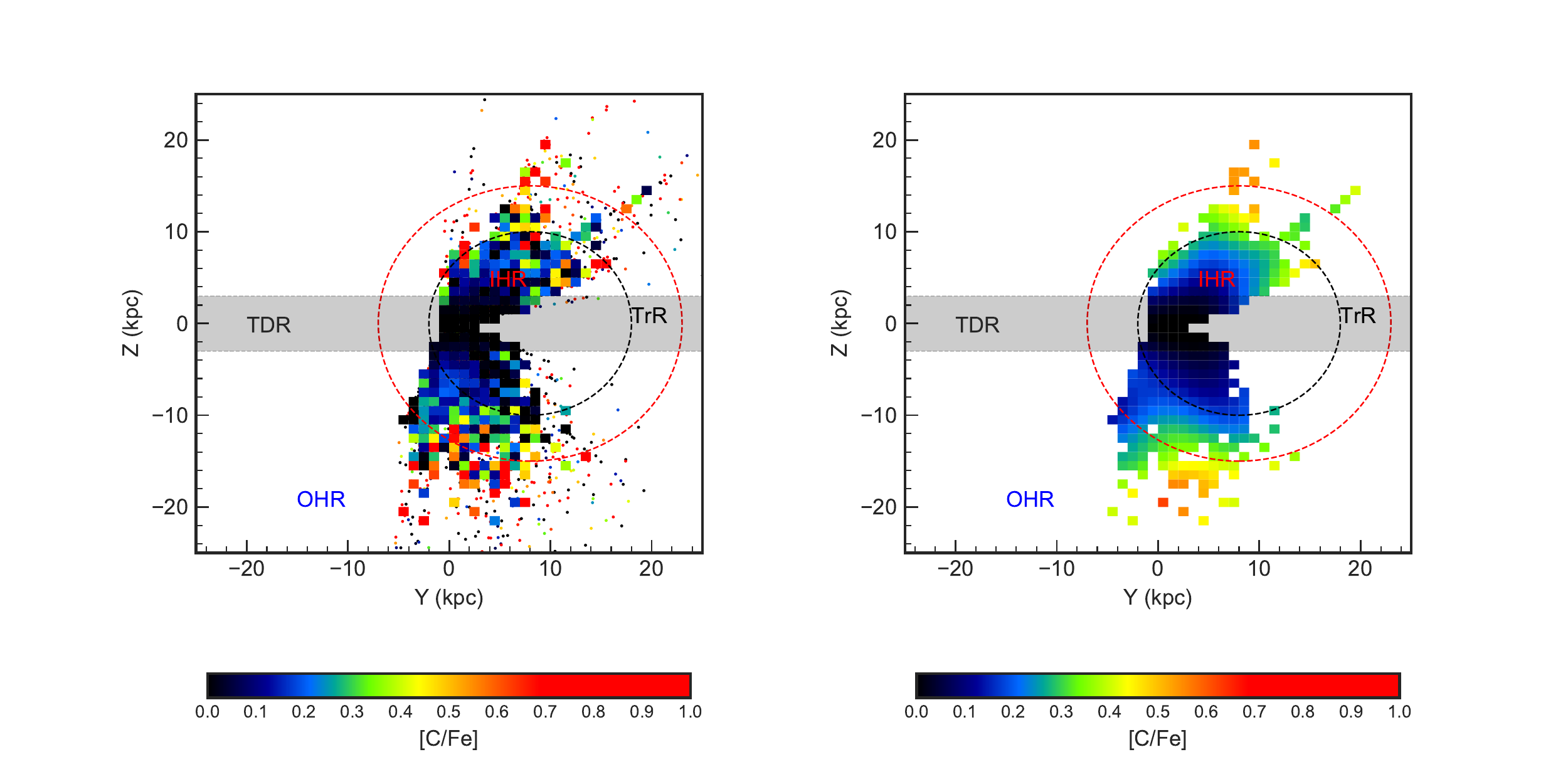}{0.9\textwidth}{(a) Carbonicity map}}
\gridline{\fig{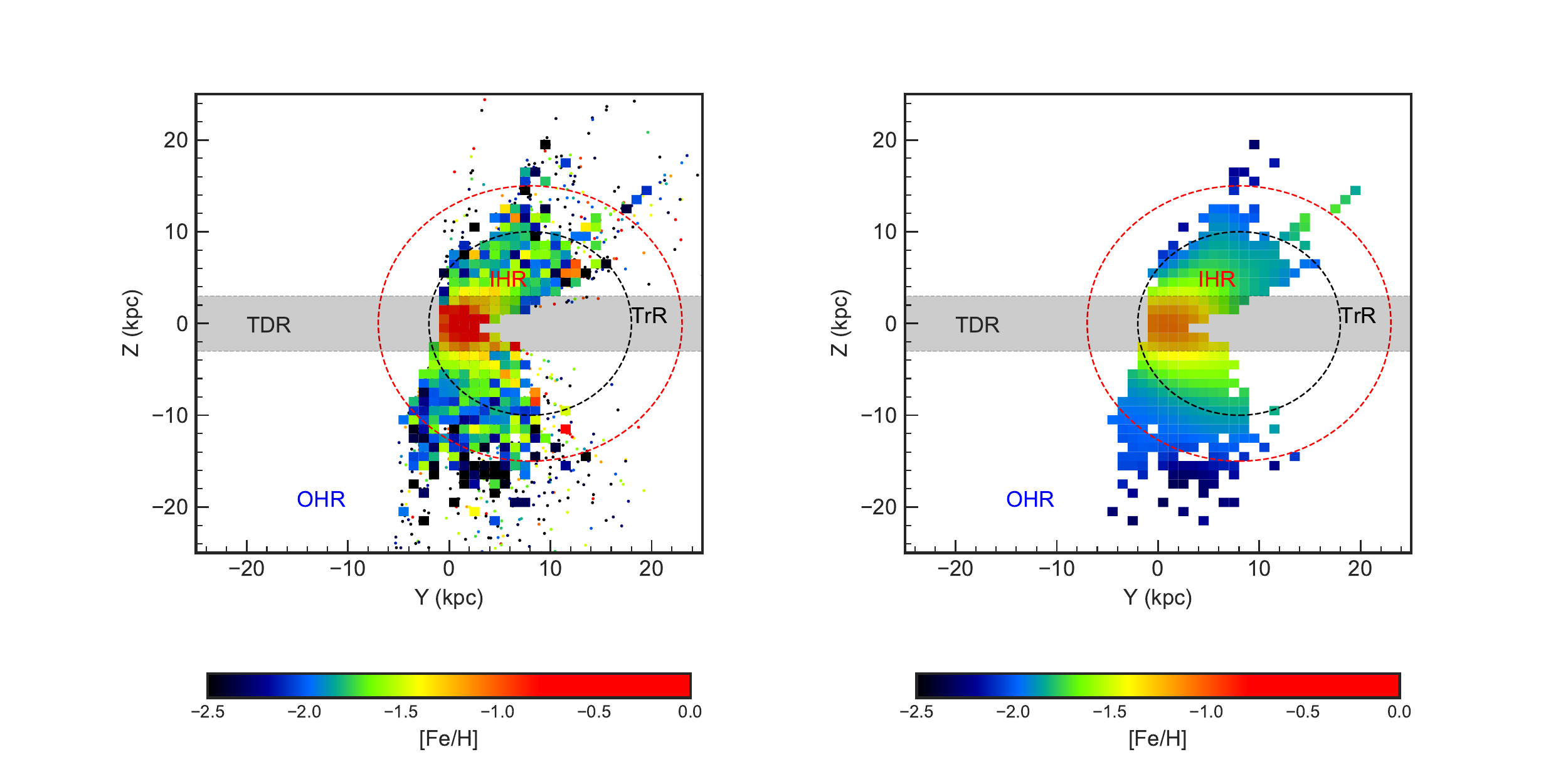}{0.9\textwidth}{(b) Metallicity map}}
\caption{Cartographic maps of carbonicity (upper panels) and metallicity
(lower panels) projected onto the 
Galactocentric Y-Z plane.  The symbols and color scales are the same as in Figure~\ref{fmapsxz}.
\label{fmapsyz}}
\end{figure*}

\section{Galactic Cartography of Carbonicity and Metallicity}\label{sec:cartography}

\begin{figure*}[th!]
\gridline{\fig{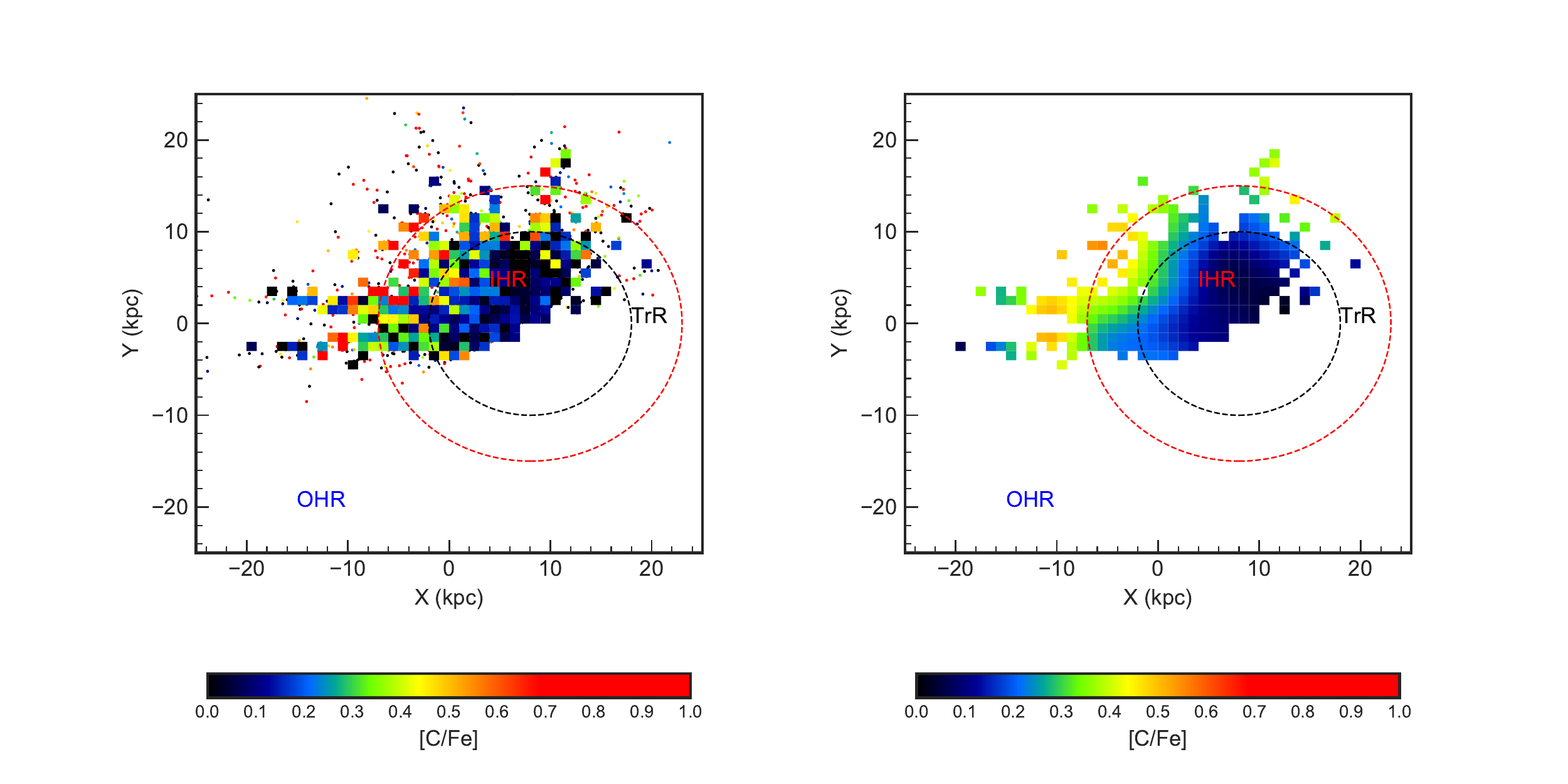}{0.9\textwidth}{(a) Carbonicity map}}
\gridline{\fig{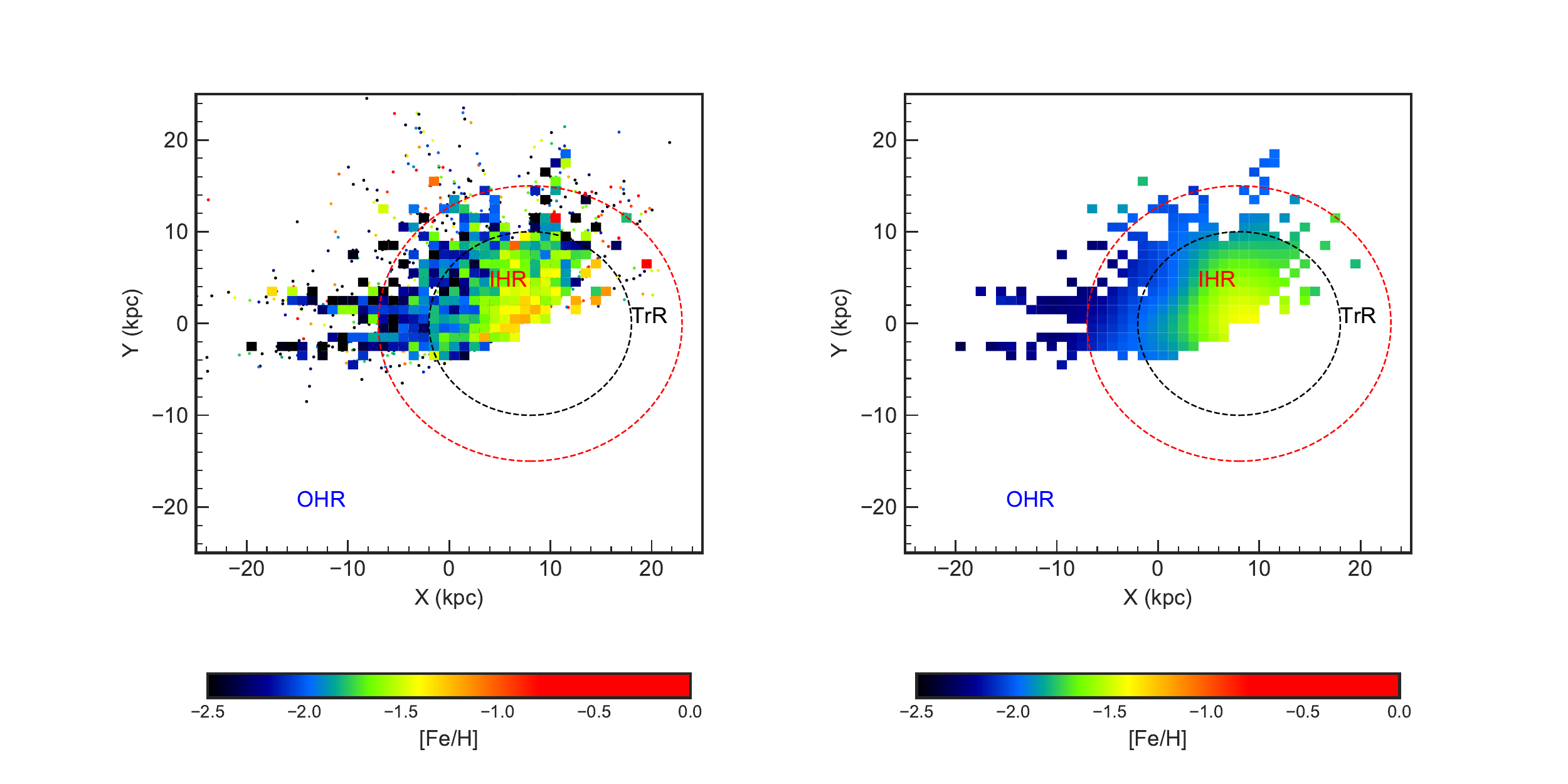}{0.9\textwidth}{(b) Metallicity map}}
\caption{Cartographic maps of carbonicity (upper panels) and metallicity
(lower panels) projected onto the Galactocentric X-Y plane.  
The symbols and color scales are the same as in Figure \ref{fmapsxz}. Stars
in the thick-disk region (TDR: |Z| $ \leq 3$ kpc) are removed from this plot.
\label{fmapsxy}}
\end{figure*}

Figures \ref{fmapsxz}-\ref{fmapsxy} are Galactocentric cartographic maps
(projected onto the X-Z, Y-Z, and X-Y planes, respectively, in right-handed
rectangular Galactocentric coordinates, having positive X towards the Galactic anti-center) of carbonicity (upper panels)
and metallicity (lower panels). The left panels in each figure show the distribution of stars in a square grid of (0.5 kpc $\times$ 0.5 kpc) pixels. The filled squares have pixels with at least three stars, and the filled dots indicate pixels with two or one star. The right panels of each figure show the stellar distribution in the pixel grid, smoothed with a two-pixel Gaussian kernel. The color bar under each panel 
corresponds to the median values of the [C/Fe] and [Fe/H] values shown in the maps.

\subsection{Galactic Components Based on Carbonicity}\label{sec:GC}

In order to identify individual Galactic components, and to consider the
nature of the [C/Fe] and [Fe/H] distributions within them, we follow the
approach of \citet{lee2017}, who made use of carbonicity 
to make these assignments (rather than metallicity or kinematics), with
a few adjustments. For example, Lee et al. constructed dividing lines
based on the cylindrical Galactocentric R-|Z| plane, whereas, in this work, we
mapped the distributions of [C/Fe] and [Fe/H] projected onto the
three rectangular Galactocentric planes (X-Z, Y-Z, and X-Y). 

Our divisions based on carbonicity, shown in Figures
\ref{fmapsxz}--\ref{fmapsxy}, are obtained as follows. Both the dashed
circles are centered around the Solar Neighborhood, at R = 8\,kpc, Z = 0 kpc (\citealt{bovy2015}; for convenience, we used Z = 0 kpc rather than Z = 0.025 kpc), and the inner black and outer red circles have radii of 10\,kpc and 15\,kpc, respectively. The inner and outer circles correspond to the median value of [C/Fe] $\sim +0.2$ and [C/Fe] $ \sim +0.4$ to +0.5, respectively. (We note that Lee et al. used [C/Fe] $\sim$ +0.4 and +0.6 for separating the halos, resulting in dividing circles located at R $\sim$ 8\,kpc and 10\,kpc respectively. These differences with respect to \citet{lee2017} are purely data-driven; Lee et al. only made use of SDSS/SEGUE main sequence turn-off stars, while we employed stars over a wider range of luminosity in the AEGIS data, including more distant giants.) We then divided each map into four Galactic components, a
thick-disk region (TDR; gray-shaded area, |Z| $\leq 3$\,kpc, roughly three thick-disk scale heights above the Galactic plane), an inner-halo region (IHR; |Z| $> 3$\,kpc and inside
the black-dashed circle), a transitional region (TrR; |Z| $ > 3$\,kpc
and between the dashed circles), and an outer-halo region (OHR; |Z| $ >
3$\,kpc and outside the red-dashed circle). The individual components
are labeled in the figures. Each region is dominated by stars in the
indicated component population, yet still suffers contamination from
other components, as described in detail in the next subsection. We note
that, although the divisions we made are based on carbonicity, they are
similar to previously suggested dividing lines for the IHR, TrR, and OHR
based on either kinematics or metallicity \citep[e.g., ][and references
therein]{carollo2007,carollo2010,dejong2010,tissera2014, bland-hawthorn2016}.\\

\subsection{Metallicity Distributions in the Galactic Components}

The Galactic components identified based on their carbonicity levels
clearly correspond to different mean metallicities. As seen in the
metallicity maps, the IHR exhibits a value near [Fe/H] $\sim
-$1.5, while the OHR exhibits $-2.5 < $ [Fe/H] $ < -2.0 $, on average.
Figure \ref{fdf} shows the MDFs, the carbonicity distribution functions, and the absolute carbon abundance distributions in the left, middle, and right
panels, respectively. From top to bottom, the columns of panels indicate the TDR, IHR,
TrR, and OHR regions, respectively. The gray-shaded histogram indicates
all stars in the sample. The magenta and green histograms represent
dwarf/turn-off (D/TO) stars and subgiant/giant (SG/G) stars,
respectively. We note that there are a small fraction of D/TO stars in the OHR, which are likely to be spurious. We left out these D/TO stars in the green histogram representing the OHR in the bottom panels. The green-, blue-, and red-dashed vertical lines represent
the mean metallicity ([Fe/H] = $-$0.6, $-$1.6, and $-$2.2) of the thick-disk, inner- halo, and outer-halo populations, respectively (see, e.g., \citealt{carollo2010} and \citealt{an2013}).  Details of the metallicity distributions for the stars in 
each region are provided below.

\begin{enumerate}
\item Thick-Disk Region (TDR) -- The peak metallicity of the stars in this region 
is located at [Fe/H] $\sim -$0.7, and this region is likely 
dominated by stars of the thick-disk population. A large fraction
($\sim$74\%) of the population in the sample consists of dwarfs and
turn-off stars, unlike the other three components, which are
predominantly sub-giants or giants. The peak metallicity of the D/TO
MDF is [Fe/H]$ \sim -$0.7, commensurate with many studies \citep{carollo2010,an2013}, while the SG/G stars exhibit a peak metallicity at [Fe/H]$\sim -$1.4, and may suffer from contamination from the inner-halo population (IHP). 
\item Inner-Halo Region (IHR)-- The dominant population is comprised of sub-giants and giants with a peak metallicity at [Fe/H] $ = -$1.6, which is clearly distinct
from the thick disk.
\item Transitional Region (TrR) -- Sub-giants and giants are dominant in this region 
as well, and they reflect a mixture of the IHP at [Fe/H] $ = -$1.6
and the outer-halo population (OHP) at [Fe/H]$ = -$2.2.
\item Outer-Halo Region (OHR)-- The sub-giants and giants that dominate this 
region include contributions from both the IHP and OHP. The tail of
lower-metallicity stars in the MDF is clearly stronger than for the other
regions.
\end{enumerate}

The enumerated results above for the stellar populations represented in the Galactic
components are quite similar to those obtained from the main-sequence
turn-off stars from SDSS studied by \citet{lee2017}. As in that work, it
is interesting to see that the Galactic components identified by
carbonicity cuts provide independent evidence for the existence of the
Galactic components in metallicity space.

\subsection{Distribution of Carbon in the Galactic Components}

\begin{figure*}%[th!]
\centering
\includegraphics[scale=.65]{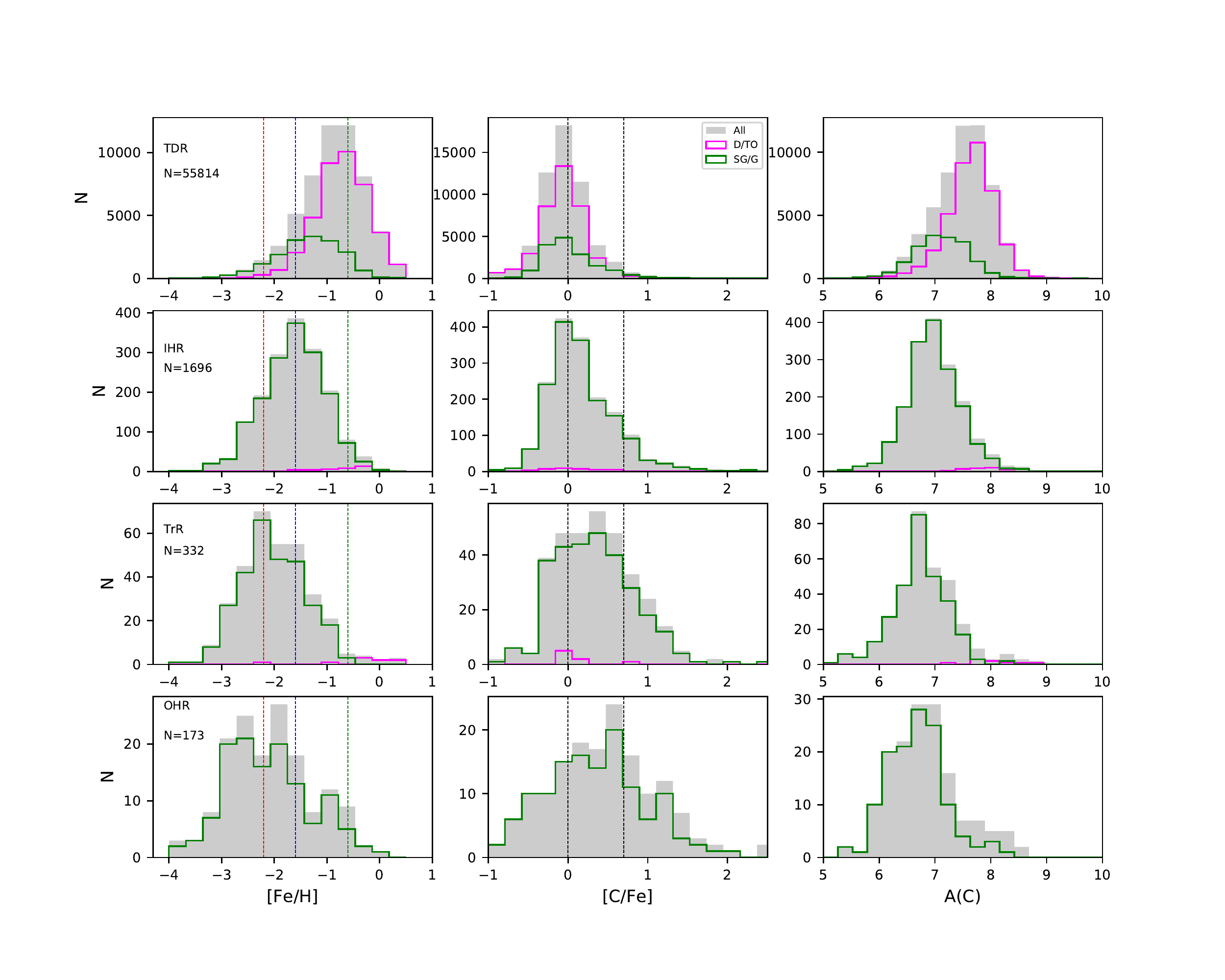}
\caption{ Distributions of metallicity (left panels), carbonicity (middle panels),
and absolute carbon abundance (right panels). The carbonicity estimates were corrected for SG/G stars by adopting the evolution-correction calculation of \citet{placco2014c}. The top to bottom panels indicate the TDR, IHR, TrR, and OHR, respectively. The gray-shaded,
magenta-solid, and green-solid histograms represent all stars including FHB stars, D/TO stars only, and
SG/G stars only, respectively. The total number of stars in each Galactic component is reported underneath each component designation in the left panels. The red-, blue-, and green-dashed vertical lines represent the mean metallicity of the outer halo, inner halo, and thick disk, respectively \citep{carollo2010}. The black-dashed lines in
the middle panels indicate [C/Fe] = 0.0 and [C/Fe] = +0.7. 
\label{fdf}}
\end{figure*}

The middle panels of Figure \ref{fdf} indicate that the level of
carbonicity increases with decreasing metallicity from the TDR to the OHR, as shown in
the cartographic maps. 
The fraction of stars with higher carbonicity increases from the TDR to the OHR as well. 

The $A$(C) distribution (shown in the right-hand panels) in each
component also shifts toward lower values from the TDR to the OHR. The dominant
population of D/TO stars in the TDR have a peak at $A$(C) $\sim $ 7.8, while the
SG/G population exhibits a peak at $A$(C) $\sim$ 7.0, a difference of
$\sim$ 0.8 dex. This can be accounted for by the difference in metallicity between the D/TO stars and the SG/G stars (both having [C/Fe]$\sim$ 0.0), due to the different sampling of the populations resulting from the higher luminosities of the SG/G stars. 

\section{Carbon-Enhanced Metal-Poor Stars in the AEGIS Sample}\label{sec:cemp}

\subsection{The CEMP Population}

We now explore the properties of the CEMP stars present in the AEGIS sample.
As seen in the middle panels of Figure~\ref{fdf}, the fraction of CEMP
stars increases from the TDR to the OHR, although only about 3\% of the
AEGIS sample are CEMP stars. The TDR has a negligible fraction of CEMP
stars, since most are D/TO stars, whose higher effective temperatures
make identification of CEMP stars difficult \citep{lee2013,placco2014c,
placco2016a}. However, moving from the IHR to the OHR reveals a
substantial increase in the fractions of CEMP stars. There are a total
of 1,691 CEMP stars in the AEGIS sample, comprising 1,109 SG/G stars, 433
D/TO stars, and 149 field horizontal-branch stars. Here we only consider the SG/G and D/TO stars, resulting in a total of 1,542 CEMP
stars for the classification and frequency analysis described below. 

\begin{figure*}[th!]
\gridline{\fig{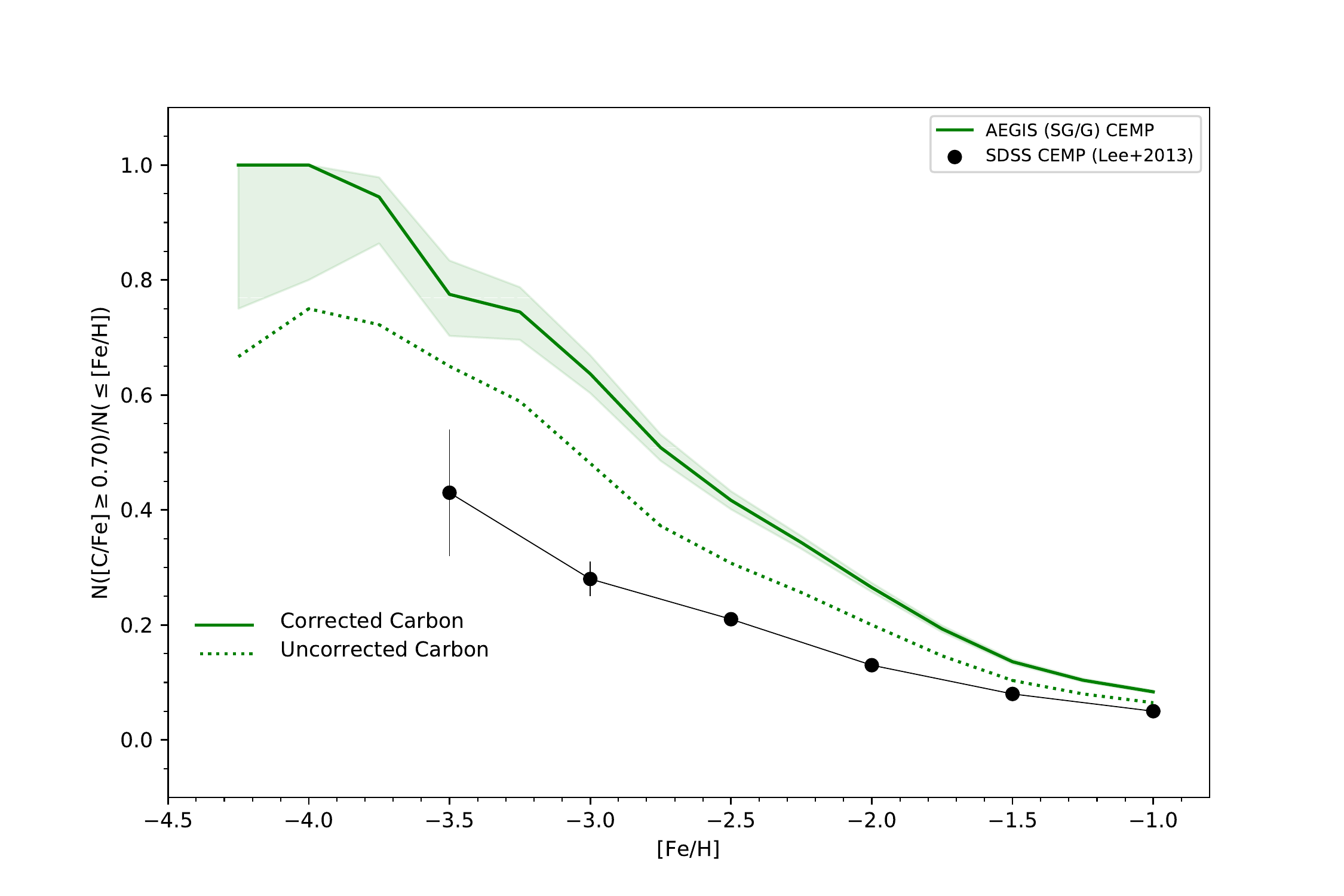}{0.5\textwidth}{(a)
Cumulative frequencies of all CEMP stars. }
\fig{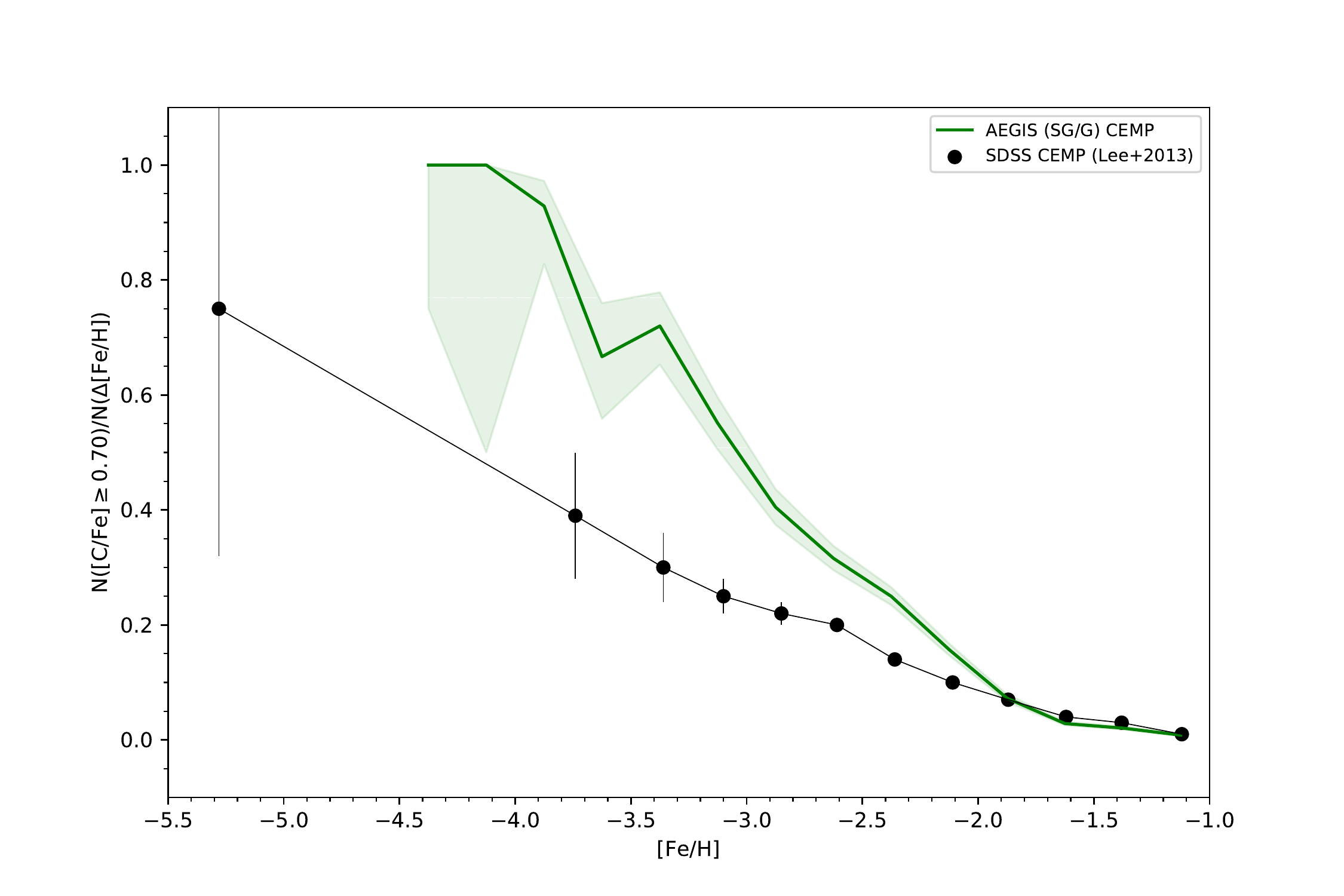}{0.5\textwidth}{(b)
Differential frequencies of all CEMP stars. }}
\gridline{\fig{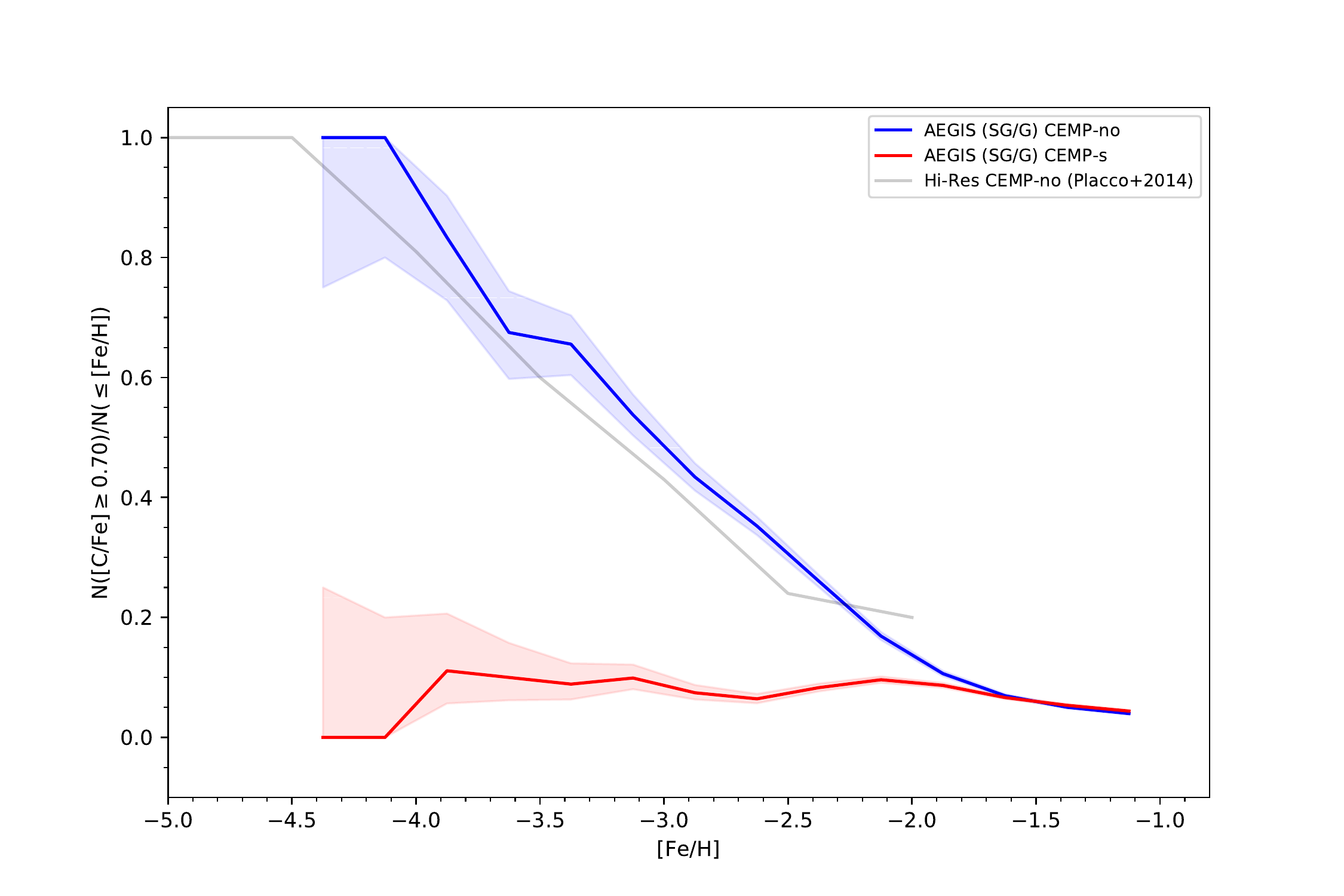}{0.5\textwidth}{(c)
Cumulative frequencies of the CEMP-no and CEMP-$s$ stars. }
\fig{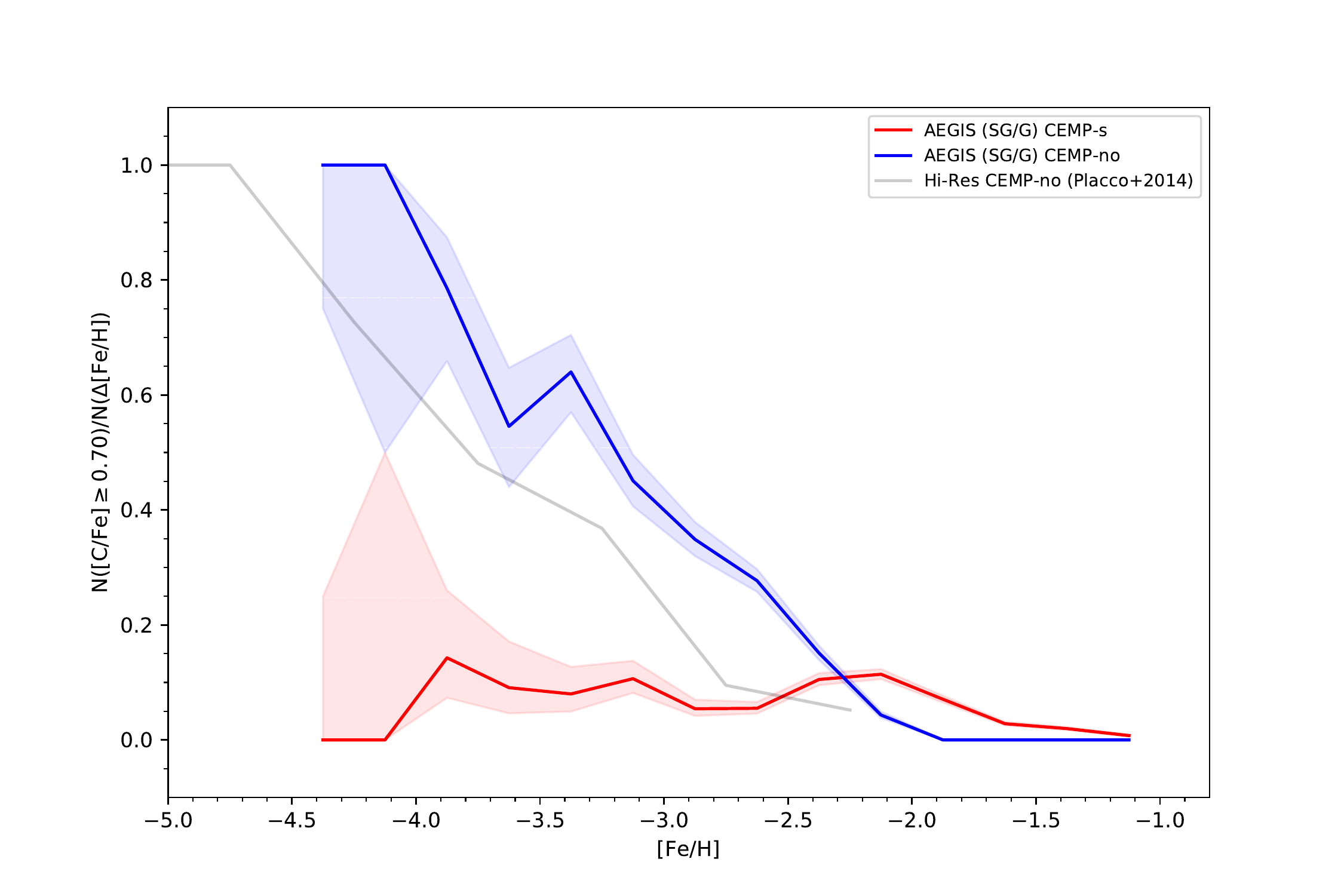}{0.5\textwidth}{(d)
Differential frequencies of the CEMP-no and CEMP-$s$ stars. }}
\caption{Panels (a) and (b) show the cumulative frequencies at a given metallicity
and the differential frequencies of CEMP SG/G stars at each metallicity bin, respectively.
Panels (c) and (d) represent the cumulative frequencies and differential frequencies of the CEMP-no and CEMP-$s$ stars, respectively. Frequencies are calculated relative to all SG/G stars. In panel (a), the green-dotted line represents the cumulative
frequency based on the observed carbon abundances, and the solid lines in all four panels show that the frequencies taking account of the carbon-abundance corrections from \citet{placco2014c}. The green-, blue-, and red-solid lines represent the frequencies of all CEMP stars regardless of their sub-classes, CEMP-no only, and CEMP-$s$ only stars, respectively. 
The lightly colored shaded areas around the solid lines indicate 1$\sigma$ CI$_W$ of individual frequencies. The black-filled circles with error
bars in panels (a) and (b) indicate the frequencies of SDSS/SEGUE CEMP stars from
\citet{lee2013}. The gray lines in panels (c) and (d) represent the high-resolution sample of CEMP-no stars from \citet{placco2014c}. Details can be found in the text.
\label{fcf}}
\end{figure*}

\subsection{CEMP Classifications} 

It is important, where possible, to distinguish between the 
sub-classes of CEMP stars, as each may correspond to a different class
of stellar progenitor(s), and explore different epochs of the assembly and 
chemical evolution of the Galaxy \citep[e.g.,][]{chansen2016}. 
Until quite recently, it was thought that such classification required 
high-resolution spectroscopy, in order
to detect the heavy elements Ba and Eu that form the basis of the
sub-class assignments as described by \citet{beers2005}.  However, based on
a large sample of CEMP stars with available high-resolution
spectroscopic classifications, \cite{yoon2016} demonstrated that CEMP-no
stars (whose surface abundances are believed to be reflect the gas from 
which they formed, \citealt{hansen2016a})
could be reasonably well-distinguished from the class of CEMP-$s$ stars
(whose surface abundances reflect a mass-transfer event from a former
AGB companion) on the basis of their absolute carbon
abundances alone, without the use of high-resolution spectroscopy. According to the Yoon et al. study, their method employing $A$(C) enabled classification of CEMP stars with $A$(C) > 7.1 as CEMP-$s$ stars and those with $A$(C) $\leq$ 7.1 as CEMP-no stars, with a success rate of $\sim$ 90\%.  In the following analysis, it is understood that the CEMP-$s$ and CEMP-no stars are classified as such through application of this approach.  

While the stars under consideration by \citet{yoon2016} were primarily
sub-giants and giants, \citet{lee2017} explored the $A$(C) distribution
of $\sim$ 100,000 main sequence turn-off (MSTO) stars from SDSS/SEGUE. They claimed that the MSTO stars require a different (higher) dividing line on absolute carbon
abundance, $A$(C) = 7.6, since in their temperature range (5600\,K $\leq$ $T_{\rm eff}$ $<$ 6700\,K) carbon molecular features become substantially weaker; identification of CEMP stars becomes increasingly difficult 
unless they have quite high carbonicity. If we use $A$(C) = 7.6 as the
dividing line for D/TO stars in the AEGIS sample, we identify 166
CEMP-no stars and 267 CEMP-$s$ stars; using $A$(C) = 7.1 for SG/G stars,
we identify 527 CEMP-no stars and 582 CEMP-$s$ stars.

\section{Frequencies of the CEMP stars}\label{sec:frequency}

The frequencies of CEMP stars as a function of metallicity provide
strong constraints on GCE \citep[e.g.,][]{kobayashi2011,cote2016,salvadori2016}, the assembly history of the Galaxy (e.g., \citealt{carollo2012,carollo2014}), and
potentially on the First Initial Mass Function \citep[FIMF; e.g.,][]{lucatello2006,tumlinson2007b, suda2013, carollo2014,yoon2016, debennassuti2017, ishigaki2018}.  They also constrain the different channels for formation of carbon-rich vs. carbon-normal stars at low metallicity \citep{norris2013b,placco2014c,chiaki2017}. 

In constructing CEMP frequencies for the AEGIS sample, we have made a
few assumptions, enumerated below.

\begin{enumerate}

\item Determining reliable chemical abundances of cool, strongly
carbon-enhanced, low-metallicity stars is very challenging, since the
strong molecular carbon bands can significantly depress the 
continuum level. We thus limit our consideration to stars with effective
temperatures $T_{\rm eff} > 4000$~K.

\item Due to the difficulty of identifying the carbon enhancement for 
warmer stars, we have limited our consideration of frequencies to the
SG/G stars in the AEGIS sample.  We note from the discussion above that the SG/G stars are the dominant population in the halo system (both for the inner halo and the outer halo).

\item Since we only include the SG/G stars, we have used $A$(C) = 7.1 for 
separating CEMP-no stars from CEMP-$s$ stars, as in \citet{yoon2016}. 

\end{enumerate}

Figure~\ref{fcf} shows the resulting derived frequencies, as a function of
[Fe/H], for the CEMP stars in the AEGIS sample. This figure shows four panels 
of frequencies defined as follows; (a) cumulative frequencies of all CEMP stars (regardless of their sub-class), (b) differential frequencies of all CEMP stars, 
(c) cumulative frequencies of each CEMP sub-class (CEMP-no and CEMP-$s$ plotted separately), 
and (d) differential frequencies of each CEMP sub-class. In all cases the definition [C/Fe] $\geq +0.7$ was used for identification of CEMP stars in the AEGIS sample.
The dotted line in panel (a) represents the cumulative frequency of 
all CEMP stars using the derived carbon abundances from the n-SSPP, 
which can reflect diluted abundances for evolved stars due to first dredge-up. 
The green-, blue-, and red-solid lines in Figure~\ref{fcf}
represent the results based on application of the carbon correction
procedure of \citet{placco2014c} for all CEMP, 
CEMP-no, and CEMP-$s$ stars from the AEGIS survey, respectively. The light-green, light-blue, and light-red shaded areas represent the Wilson score confidence intervals (CI$_W$;  \citealt{wilson1927})\footnote{The Wilson score approximation is used for estimating binomial proportion confidence intervals, as recommended by \citet{brown2001}. This approximation is commonly used for small sample size, n $\leq$ 40. For larger n$>$40, the Wilson and other approximations are comparable. Therefore we chose CI$_W$ for the fractions over all metallicity regimes.}. For comparison, the black circles in panel (a) represent the cumulative frequencies for stars with [C/Fe] $\geq$ +0.7 from the SDSS/SEGUE data of \citet{lee2013}. Note that Lee et al. made use of stars with 4400\,K 
$<$ $T_{\rm eff}$ $<$ 6700\,K, 
S/N $\geq$ 20.0, and all luminosity classes (D, TO, SG, and G). 
The SDSS/SEGUE differential frequencies for the CEMP stars (black circles) 
are included for comparison in panel (b). The SDSS/SEGUE frequencies 
include all classes of CEMP stars (there was no mechanism to differentiate 
sub-classes at the time), as well as all stars in the various luminosity classes. 
The bottom two panels (c) and (d) include the frequencies, calculated from the extensively compiled dataset of the CEMP-no stars, carried out with high-resolution spectroscopy \citep{placco2014c} for comparison with the AEGIS CEMP-no sample. It is clear that the CEMP frequency estimate based on the SDSS/SEGUE data is substantially lower 
than the CEMP-no star frequencies in the AEGIS sample, and higher than the CEMP-$s$ star frequencies seen in panels (c) and (d). 

We draw the following inferences from inspection of Figure~\ref{fcf}.
\begin{enumerate}

\item The difference between the green-solid line  and green-dotted line in panel (a) of Figure \ref{fcf} shows
that it is necessary to include the evolutionary corrections for carbon
dilution, as it changes the estimates by on the order of 10-30\%, depending on the metallicity.

\item The cumulative frequencies (the green lines) of the CEMP stars in panel (a) increase with 
decreasing metallicity, as has been reported by previous studies. 
However, since \citet{lee2013} 
included both D/TO and SG/G stars in their counts (denominator as well as numerator), and did not correct the carbon abundances according to evolutionary status for their frequency calculation, their
final frequencies (the black line with dots) ended up being about a factor of two smaller
than our result (the green-solid line). We attribute this result to both the uncorrected carbon abundances and the
difficulty of identifying CEMP stars (in particular for CEMP-no stars,
due to their substantially lower $A$(C) at a given [C/Fe]) for
warmer stars, effectively removing true CEMP stars from the numerator, and the addition of a substantially larger fraction of D/TO stars, relative to SG/G stars, to the denominator. 

\item As seen in panel (b) of Figure~\ref{fcf}, the differential
frequencies of the CEMP stars from the SDSS/SEGUE and the AEGIS sample
both increase with decreasing metallicity. However, as for the cumulative 
frequencies noted above, the \citet{lee2013} differential frequencies for the
SDSS/SEGUE sample are substantially lower than found for the AEGIS
sample, due to the uncorrected carbon abundances and the different luminosity 
classes that were included in the counting exercise. 

\item Both the cumulative and differential frequencies of the CEMP-no stars 
steeply increase with decreasing metallicity, as seen in panels (c) and (d).  Since there was no way to differentiate CEMP sub-classes for the SDSS data (at that time), we cannot directly compare our result with the SDSS data frequencies \citep{lee2013}. However, the frequencies based on high-resolution data for the CEMP-no stars \citep{placco2014c} 
clearly support not only our calculation of the frequencies, but also tacitly validate that the $A$(C) classification method is as effective as that of the conventional [Ba/Fe] criterion, even though there are some small differences in the fractions shown in panels (c) and (d). We also note that this consistency of the frequencies is likely to arise from the fact that the high-resolution sample predominantly comprises sub-giants and giants, unlike the SDSS data reported by \citet{lee2013}. 

\item In panels (c) and (d), our tiny sample of stars with [Fe/H]$\leq -$4.0 has a 100\% (with a 1$\sigma$ CI$_W$ of 20\%) frequency of CEMP-no stars (4 out of 4 stars in the metallicity bin; one star is a Group III star and three are Group II stars, according to the criteria of \citealt{yoon2016}). We note that there are more stars in these two groups with [Fe/H] $>-$4.0. However, there is a transitional region ($-3.5 <$ [Fe/H]$ <-2.5$ and $A$(C)$<$7.1), where both Group II and Group III stars reside; higher-resolution spectroscopy of $A$(Mg) and/or $A$(Na) is required for clear separation in this metallicity range.

\item A transition in the dominant stellar population from CEMP-$s$ stars to CEMP-no stars with decreasing metallicity is clearly seen at [Fe/H]$\sim -2.3$ in panel (d). 

\item Both the cumulative and differential frequencies of the CEMP-$s$ stars in the panels (c) and (d) are flat ($\sim$10\%) for the stars with [Fe/H] $\lesssim -2.0$, consistent with the CEMP frequency at [Fe/H] $\sim -$2.3 obtained by \citet{abate2015c} (between 7\% and 17\%). Their CEMP frequency was based on their synthetic stellar-population models (which only included binary mass-transfer origins for CEMP stars) and then compared with the observed CEMP fractions for SDSS/SEGUE stars from \citet{lee2013}. They found an inconsistent result, that their theoretical CEMP fraction was a factor of two lower than that of the observed data. The reason for this discrepancy is now made clear; at low metallicities, the CEMP-$s$ stars must be separated from the increasingly common CEMP-no stars prior to the comparison being made.

\end{enumerate}

\section{Discussion}\label{sec:discussion}

The formation of the Galaxy and its chemical-evolution history are closely interconnected. In particular, the spatial distribution of the stellar chemical elements in different regions provides information not only on the various stellar populations, but also on their natal environments, which helps to constrain their stellar progenitors.  Here we have used a new large sample of medium-resolution spectra for stars in the Southern Hemisphere, the AEGIS survey, to explore the spatial distributions of C and Fe, and consider the frequencies of CEMP stars. For the first time, we have been able to sub-classify the stars into CEMP-no and CEMP-$s$ stars, using medium-resolution spectra alone.  These results are discussed below, in the context of the dual halo model of the Milky Way.

\subsection{The Dual Halo System as Revealed by Carbonicity}

\citet{lee2017} constructed the first carbonicity maps of the Galactic halo, based on a large sample of MSTO stars from SDSS/SEGUE. Their carbonicity map indicated a clear dichotomy of the halo system in terms of the relative fractions of the two most populous CEMP sub-classes -- the low $A$(C) stars associated with CEMP-no stars were found to dominate the OHR, while the high $A$(C) stars associated with CEMP-$s$ stars dominate the IHR. This result provided support to the initial claim for this segregation made by \citet{carollo2012}, based on a much smaller sample of CEMP-no and CEMP-$s$ stars classified on the basis of available high-resolution spectroscopy.

Following the Lee at al. prescription to divide the halo system based on its distribution of carbonicity, inspection of the relative fractions of CEMP-no and CEMP-$s$ stars in the AEGIS survey (considering only the SG/G stars) revealed a similar behavior.  The IHR comprises 
47$\pm$4\% CEMP-no stars and 53$\pm$4\% CEMP-$s$ stars, the TrR comprises 64$\pm$6\% CEMP-no and 36$\pm$6\% CEMP-$s$ stars, and the OHR comprises 78$^{+6}_{-8}$\% CEMP-no and 22 $^{+6}_{-8}$\% CEMP-$s$ stars; errors in the frequencies were calculated based on the 1$\sigma$ CI$_W$. Although the fractions of the sub-classes differ somewhat from those found by Lee et al., the dominant population in each Galactic component is consistent with their result.

Both the Lee et al. results and ours can be understood in terms of our current picture of the formation of the inner- and outer-halo populations of stars (summarized in the next sub-section). The relatively more massive ($\sim 10^9$M$_\odot$) mini-halos (classical dwarf galaxy counterparts) that formed the IHP led to the production of larger fractions of CEMP-$s$ stars that dominate the IHR, while the relatively less massive ($\sim 10^6$M$_\odot$) mini-halos (ultra-faint dwarf galaxy counterparts) that were accreted to form the OHP led to larger fractions of CEMP-no stars in the OHR. The MDFs of low-mass mini-halos, on average, span a broader [Fe/H] range, with much lower metallicity tails, than more massive mini-halos, due to their  truncated star-formation history. Thus they contain more of the most metal-poor stars with [Fe/H]$<-$3.0, which are predominantly CEMP-no stars. However, massive mini-halos have more metal-rich stars, due to more prolonged star formation, thus we expect the CEMP-$s$ stars to dominate over the CEMP-no stars in these environments \citep{salvadori2015,salvadori2016}. 

\subsection{Galactic Formation History as Revealed by Metallicity}\label{sec:formation}

\citet[ELS;][]{eggen1962} proposed a rapid monolithic collapse model of the Galactic halo, which was later challenged by
\citet[SZ;][]{searle1978}, who claimed that the formation
of the halo was due to the accretion of ``protogalactic fragments" that
continued to fall into the Galaxy after formation of the central region was complete.  
Aspects of both the ELS model and the SZ model were reflected in subsequent observational work and simulations, which converged to favor the halo accretion model \citep[e.g.,][]{samland2003, steinmetz1995, chiba2000, bekki2001, brook2003, bullock2005, diemand2005, zolotov2009}. 

More recent studies, using large samples of stars from SDSS/SEGUE, were able to demonstrate the existence of at least 
two distinct stellar halos -- the inner halo and the outer halo -- 
based on the spatial distributions of metal-poor stars, and correlations between with kinematics and metallicity \citep[e.g.,][and references therein]{bekki2001, carollo2007,delucia2008, carollo2010, dejong2010, beers2012, xu2018}.
The flattened IHR is dominated by contributions from the IHP at distances $\leq$ 10-15\,kpc, while the more spherical OHR is dominated by contributions from the OHP beyond 15-20\,kpc.  As supported by more recent numerical simulations \citep[e.g.,][]{amorisco2017,starkenburg2017},  
the outer halo is likely to have formed via essentially dissipationless accretion of low-mass mini-halos, whereas the inner halo formed via dissipative merging 
between more-massive mini-halos. 

The dual halo components selected by applying the carbonicity cuts in both the Lee et al. and the AEGIS sample are clearly well-represented as distinct peaks in the MDFs of the stars in the IHR and OHR, at [Fe/H] $\sim -$ 1.6 and $-$ 2.2, commensurate with the results of previous studies based on the density distribution and kinematics of halo stars \citep[e.g.,][]{carollo2010, beers2012, an2013, an2015, das2016}. 

\subsection{Constraints from the Frequencies of CEMP Stars}

The frequencies of CEMP stars have been considered in a number of  previous studies based on a variety of samples \citep{cohen2005, lucatello2006, frebel2006b, carollo2012,lee2013, placco2014c, beers2017}, all of which concluded that the cumulative frequency of CEMP stars strongly increases with decreasing metallicity. However, our present analysis differs in that we have limited our calculations to consider only SG/G stars, due to the recognition that including the (generally warmer) D/TO stars leads to a distortion of the true frequencies.  The combined effects of the difficulty of detecting carbon enhancement for warmer stars and the fact that there exist several orders of magnitude difference in the $A$(C) for CEMP Group I stars vs. CEMP Group II and III stars seen in the Yoon-Beers diagram confounds the naive calculation of frequencies which ignore them.  The net result is to lower (by up to a factor of two) the reported frequencies of CEMP stars from their correct values.
Secondly, we have reported here, for the first time, the cumulative and differential 
frequencies of individual CEMP sub-classes (CEMP-no and CEMP-$s$). Because these sub-classes have different astrophysical origins, it is necessary to distinguish them in 
order to place reliable constraints on chemical-evolution and stellar 
population-synthesis modeling. 

Important implications can be drawn from inspection of the individual frequencies for CEMP-no and CEMP-$s$ stars, summarized below: 
\begin{enumerate}
\item The CEMP-no fraction  in the extremely low-metallicity regime ([Fe/H] $<-3.0$) is sensitive to the FIMF and to the yields of the first enrichment sources, since these stars are expected to be bona-fide second-generation stars \citep[e.g.,][and reference therein]{placco2015,hansen2016a,placco2016b,yoon2016,sharma2017}.

\item The differential and cumulative frequencies of the CEMP-$s$ stars, when considered alone, are substantially lower than the frequencies of all CEMP stars. The discrepancy between the observed frequencies of SDSS CEMP stars \citep{lee2013} at low metallicity (when considered as a single population) with the predicted frequencies from population-synthesis models that only included binary mass-transfer origins \citep{abate2015c} has now been removed.  
The reason can be explained as follows. \citet{lee2013} did not have a method to separate the CEMP-$s$ stars from the CEMP-no stars at the time, so their estimated CEMP fraction included {\it both} sub-classes. However, most CEMP-$s$ stars have a binary mass-transfer origin, thus the proper comparison should be between the Abate et al. prediction and and the observed CEMP-$s$ fraction, which has been carried out in this work.

\item  The apparently constant differential fraction ($\sim 10$\%) of the CEMP-$s$ stars at $- 4.0 \,\lesssim$ [Fe/H] $\lesssim -$ 2.0 suggests that metallicity does not have a significant influence on the operation of the $s$-process, on the formation of low-mass binaries, or on their initial separation, all of which might have impacted the observational result. 
According to \citet{yoon2016}, there exist a substantial number of CEMP-$s$ stars at [Fe/H] $< -$3.0 -- there are even two CEMP-$s$ stars known with [Fe/H] $< -$3.5: CS~22960-053 with [Fe/H] $= -$3.64 \citep{roederer2014} and HE~0002-1037 with [Fe/H] $= -$3.75 \citep{hansen2016b}. The apparent cut-off metallicity at [Fe/H]$\sim -$3.8 may indicate that the emergence of AGB stars was delayed to accommodate the evolutionary timescales for intermediate-mass stars, but the numbers are still too small be to clear on this point.

For completeness, we note that spinstar production of $s$-process elements at extremely low metallicity, which would not necessarily require  a mass-transfer event, has been suggested by several authors
\citep[e.g.,][]{frischknecht2016, choplin2017} to account for the possible non-binary nature of several CEMP-$s$ stars reported by \citet{hansen2016b}. Larger samples of CEMP-$s$ stars with [Fe/H] $< -3.0$ with available high-resolution spectroscopic analyses, as well as more extensive radial-velocity monitoring, are required to evaluate these predictions. 

\item The transition of the dominant CEMP sub-class from CEMP-no stars to the CEMP-$s$ stars appears at [Fe/H] $\sim-2.3$. This can be interpreted as the transition from a FIMF (favoring more-massive-stars) to the current IMF (favoring low- and intermediate-mass stars), as discussed previously by \citet{suda2011}, \citet{suda2013}, \citet{yamada2013}, and \citet{lee2014}.

Interestingly, the fractions of the CEMP-no and CEMP-$s$ stars we obtain over the metallicity range considered in our sample ($-4.0 <$ [Fe/H] $<-$1.0) are roughly similar, in contrast to  previous suggestions that the CEMP-$s$ stars are the dominant population \citep[e.g.,][]{aoki2007}. This discrepancy likely arises due to the low $A$(C) associated with the CEMP-no stars, so that they were not recognized as CEMP stars at the main-sequence turnoff, unlike the high-$A$(C) CEMP-$s$ stars.

\end{enumerate}

\section{Summary and Future Work}\label{sec:future}
We have explored the AEGIS sample, an extensive spectroscopic data set ($\sim$58,000 stars) 
in the Southern Hemisphere, to study the origin and formation history of 
the Galactic halo, and its chemical evolution, by considering the spatial 
distributions of [C/Fe] and [Fe/H], the stellar populations, and CEMP-star 
frequencies.  We have confirmed that carbonicity increases and metallicity 
increases with distance, from the IHR to the OHR. Based on the CEMP 
population in the halo systems present in the AEGIS sample, we also 
confirm the previous results that the CEMP-$s$ stars are dominant in the 
IHR, while the CEMP-no stars dominate the OHR. Both the cumulative 
and differential frequencies of CEMP stars increase with decreasing 
metallicity. 

For the first time, we calculated the separate frequencies of 
CEMP-no and CEMP-$s$ stars, based on medium-resolution spectroscopy alone, making use of their characteristically different $A$(C) values, as described by \citet{yoon2016}. The frequencies of the CEMP-no stars are consistent with the result obtained from the extensive compilation of high-resolution data for CEMP-no stars explored by \citet{placco2014c}. The frequencies of the 
the CEMP-$s$ stars are almost constant with declining metallicity, at about 10\%, consistent with the result of  
\citealt{abate2015c} from population-synthesis modeling assuming only binary mass-tranfer origins for CEMP stars.  

To complete this effort, 
we are planning to re-calculate the frequencies of CEMP-no and CEMP-$s$ stars 
from SDSS/SEGUE data based on sub-giants and giants alone, and carry out kinematic analyses of CEMP-no/CEMP-$s$ 
stars from AEGIS, SDSS/SEGUE, and RAVE, in particular with more accurate distances and proper motions from Gaia DR2. Comparison of these observations with the predicted frequencies of CEMP-no and CEMP-$s$ stars as a function of metallicity, and the expected morphology of the $A$(C) vs. [Fe/H] diagram of \citet{yoon2016}, based on different input parameters for cosmological hydrodynamical simulations (as recently explored by \citealt{sharma2017} and \citealt{hartwig2018}), could provide powerful new constraints on the nature of star formation and chemical evolution in the early universe. In the near future, we expect to be able to identify numerous CEMP Group II and III stars, based on the morphology of the $A$(C) vs. [Fe/H] space, and  advance our understanding about the origin and nature of the  first generations of stars in the universe.

\acknowledgments
J.Y., T.C.B., S.D.,and V.M.P. acknowledge partial support
from grant PHY 14-30152; Physics Frontier Center/JINA Center for the
Evolution of the Elements (JINA-CEE), awarded by the US National Science
Foundation.Y.S.L. acknowledges support from the National Research Foundation of Korea grant funded by the Ministry of Science and ICT (No.2017R1A5A1070354, NRF-2015R1C1A1A02036658, and NRF-2018R1A2B6003961).  
This research was supported in part by the Australian Research Council through Discovery Project grants DP0878137 (Lead: B. P. Schmidt) and DP12010237 (Lead: G. S. Da Costa).  M.S. is supported by an STFC postdoctoral fellowship at the ICC, Durham. 

This research also made use of NASA's Astrophysics Data System, 
the SIMBAD astronomical database, operated at CDS, Strasbourg, France. 
This work also made extensive use of python, astropy \citep{astropy2013}, galpy\footnote{\url{http://github.com/jobovy/galpy}}\citep{bovy2015}, 
matplotlib \citep{hunter2007}, numpy \citep{numpy} and scipy \citep{scipy2001}. 

\software{astropy \citep{astropy2013}, galpy \citep{bovy2015}, matplotlib \citep{hunter2007}, numpy \citep{numpy}, scipy \citep{scipy2001} }

\appendix
\begin{figure*}[thb!]
\centering
\includegraphics[scale=0.6]{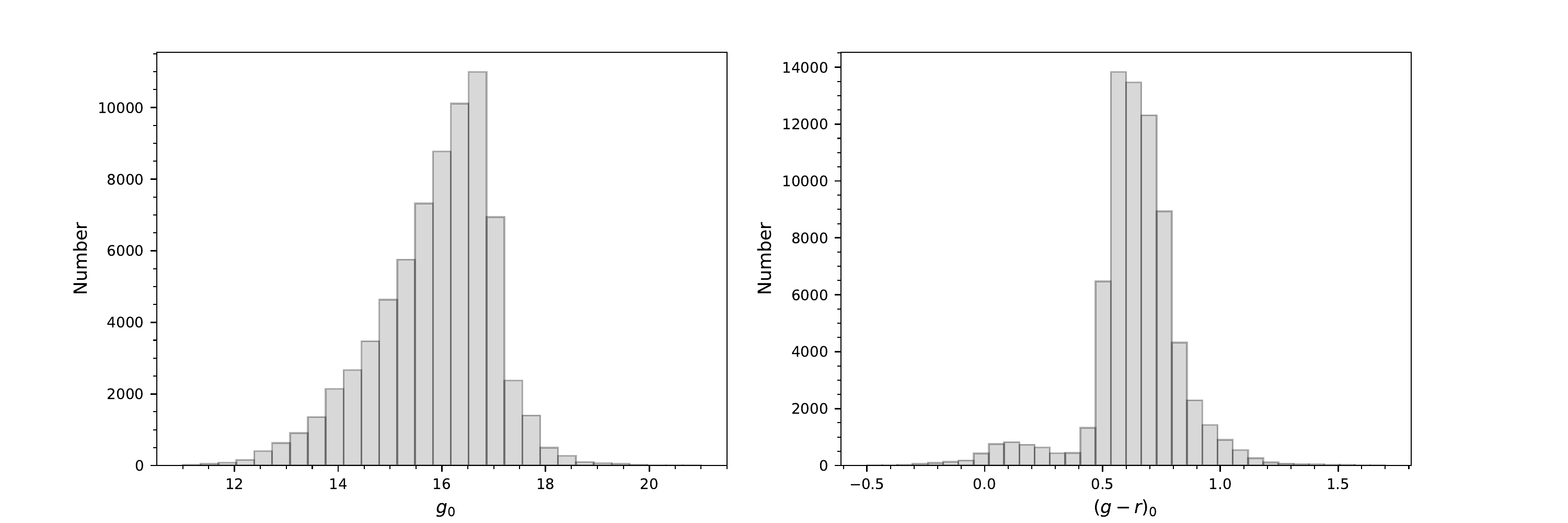}
\caption{Distributions of $g_0$ and $(g-r)_0$ \label{fmag}}
\end{figure*}

\begin{figure*}[thb!]
\centering
\includegraphics[scale=0.6]{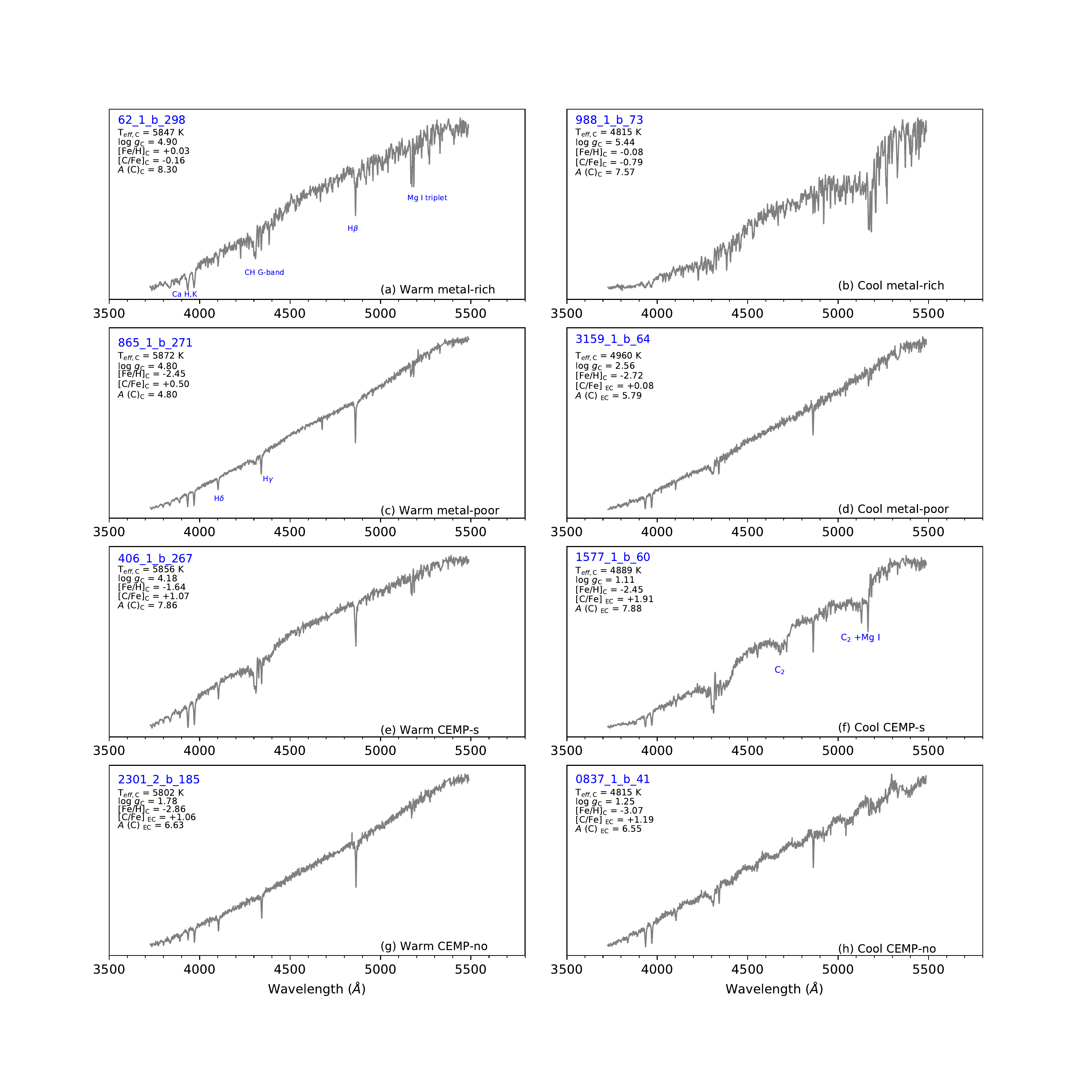}
\caption{Example AEGIS spectra. Upper row: Spectra of relatively warm (left panel) and cool (right panel), metal-rich, carbon-normal stars.  Second row: Spectra of relatively warm (left panel) and cool (right panel), metal-poor, carbon-normal stars.  Third row: Spectra of relatively warm (left panel) and cool (right panel) CEMP stars classified as CEMP-$s$, based on the $A$(C) criterion.  Bottom row: Spectra of relatively warm (left panel) and cool (right panel) CEMP stars classified as CEMP-no, based on the $A$(C) criterion.  
\label{fspectra}}
\end{figure*}

Here we summarize the nature of the AEGIS sample, provide examples of the spectra obtained, and describe the analysis procedures used for the derivation of the atmospheric parameters ($T_{\rm eff}$, $\log g$, and [Fe/H]), as well as the carbon-to-iron ratio (carbonicity), [C/Fe].  We also summarize the procedures used for the distance estimates employed.\\

\section{The AEGIS Sample}

The original photometry is obtained from commissioning era SkyMapper observations \citep{wolf2018}. Transformations from the observed $g$ magnitudes and $g-r$ colors (obtained from a calibration of $g-i$ colors to $g-r$ from APASS \citep{henden2015,wolf2018} were applied.  Reddening estimates were taken from \citet{schlegel1998}. Figure \ref{fmag} shows the distribution of $g_0$ and $(g-r)_0$ for the stars in the AEGIS sample.  The brightest stars reach $g_0 \sim 11.5$, while the faint limit of stars with available spectroscopy is $g_0 \sim 20$ but the vast majority of are brighter than $g_0 \approx$ 18.

\begin{figure*}[thb!]
\centering
\includegraphics[scale=0.5]{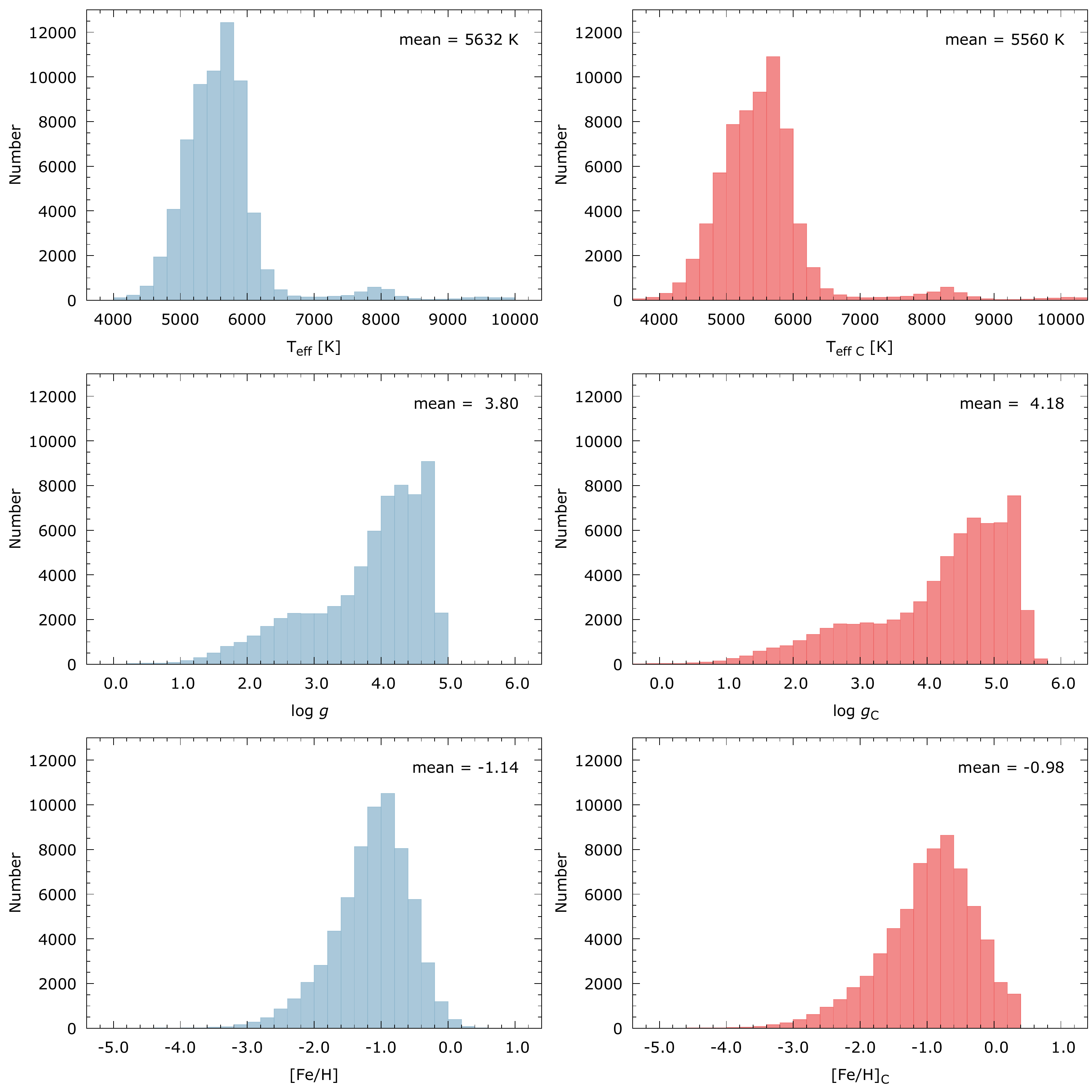}
\caption{Distributions of derived atmospheric parameters for the AEGIS sample.  The left-hand panels correspond to the n-SSPP estimates of $T_{\rm eff}$, $\log g$, and [Fe/H], from top to bottom, respectively.  The right-hand panels show the distributions of these estimates after transformation to a high-resolution spectroscopic scale ($T_{\rm eff\,C}$, $\log g_{\rm C}$, and [Fe/H]$_{\rm C}$ (see text). \label{fparams}}
\end{figure*}

The typical signal-to-noise of the blue spectra was $S/N \sim 50-55$, averaged over the full spectrum, which is sufficient to obtain atmospheric-parameter, carbonicity, and [$\alpha$/Fe] estimates. Figure~\ref{fspectra} provides examples of typical 
medium-resolution $R\sim 1,300$ spectra for the AEGIS program stars, obtained with the blue arm of the AAOmega spectrograph,  The panels in the left-hand column correspond to warmer stars ($T_{\rm eff} \sim 5800$\,K), while the right-hand panels are cooler stars ($T_{\rm eff} \sim 4800$\,K).  The upper two rows of panels are relatively metal-rich and metal-poor carbon-normal stars, according to their estimates of [C/Fe]$_{\rm C}$, (indicating a transformation to a high-resolution scale) or [C/Fe]$_{\rm EC}$ (indicating an additional application of the evolutionary carbon correction described by \citealt{placco2014c}, which only applies to giant stars; see below) respectively. The lower two rows are relatively metal-rich and metal-poor CEMP-$s$ ($A(C)_{\rm C}$ or $A$(C)$_{\rm EC}$ $> 7.1$) and CEMP-no stars ($A(C)_{\rm C}$ or $A$(C)$_{\rm EC}$ $< 7.1$), respectively.  The final derived parameters are shown for each star in the legends, and prominent spectral features are labeled.

\section{Derivation of Atmospheric Parameters and [C/Fe]}

Estimates of the stellar atmospheric parameters ($T_{\rm eff}$, $\log g$,
and [Fe/H]), and [C/Fe] were determined by employing the n-SSPP
pipeline \citep{beers2014,beers2017}, which is a ``non-SEGUE'' version of the
SEGUE Stellar Parameter Pipeline \citep[SSPP; see][for a detailed description of the procedures and calibrations used]{lee2008a,lee2008b,allende2008,smolinski2011,lee2011,lee2013}. 
The n-SSPP employs both spectroscopic and photometric ($V_0, (B-V)_0$, here, obtained from $g_0$ and $(g-r)_0$ using the transformations of \citealt{zhao2006}, and $J_0$ and $(J-K)_0$ from 2MASS; \citealt{skrutskie2006}) information as inputs, in order to make a series of estimates for each stellar
parameter. Then, using $\chi^2$-minimization matching techniques within dense grids of synthetic spectra, and averaging with other techniques as available (depending on the wavelength range of
the input spectra; see Table 5 of \citealt{lee2008a}), the best set of values is adopted. For application to the AEGIS data, the internal errors for the stellar parameters are typically 125\,K for $T_{\rm eff}$, 0.25 dex for $\log g$, and 0.20 dex for [Fe/H] and [C/Fe].

Final corrections were applied to the n-SSPP-derived parameters to match a high-resolution spectroscopic scale, as described in \citet{beers2014}. For spectra with the quality of AEGIS data, the external precision of the n-SSPP estimates of $T_{\rm eff}$, log\,$g$, [Fe/H], and [C/Fe] is on the order of 150\,K, 0.35 dex, and 0.25 dex, and 0.25 dex, respectively \citep{beers2014,beers2017}. As noted above, for giants, we also made corrections to the carbon-abundance estimates in order to restore the original amount of carbon in a star's atmosphere prior to dredge-up on the giant branch during stellar evolution \citep{placco2014c}. 

\begin{figure*}[thb!]
\centering
\includegraphics[scale=0.6]{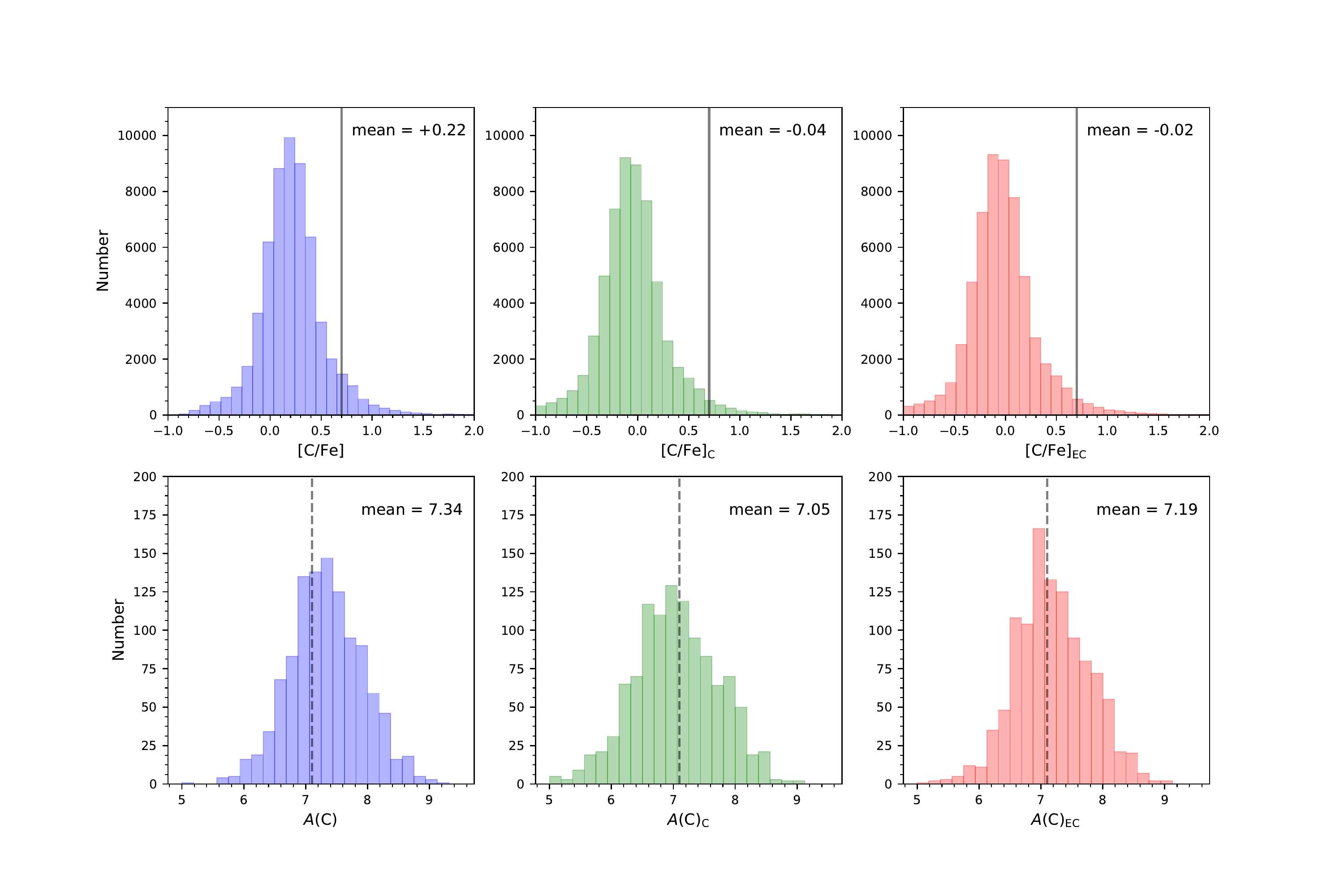}
\caption{Distributions of derived carbon-abundance estimates for the AEGIS sample. The upper row shows the n-SSPP estimates of [C/Fe], [C/Fe]$_{\rm C}$, after transformation to a high-resolution spectroscopic scale, and [C/Fe]$_{\rm EC}$, after an additional correction of the giants for the dilution of carbon during first dredge-up, from left to right, respectively. The gray vertical solid line indicates the division of carbon-normal stars from CEMP stars at [C/Fe] = +0.7. The lower row shows the corresponding distribution of absolute carbon abundances $A$(C), $A$(C)$_{\rm C}$, and $A$(C)$_{\rm EC}$, from left to right, respectively (see text). The gray vertical dashed line indicates the division on $A$(C) used by \citet{yoon2016} to differentiate CEMP-$s$ stars from CEMP-no stars at $A$(C) = 7.1.  Note that only the CEMP SG/G stars from AEGIS are shown for the bottom panels (see text).\label{fcarbon}}   
\end{figure*}

Although the n-SSPP also obtains estimates of [$\alpha$/Fe], they are not employed in the present study.  Also, in this work we only consider stars with $T_{\rm eff\,C} >$ 4000\,K, due to the difficulty of obtaining reliable parameters for these spectra in the presence of strong molecular bands. There were a total of 58,029 stars with suitable-quality spectra required to derive reliable stellar parameter estimates.

\begin{figure*}[thb!]
\centering
\includegraphics[scale=0.6]{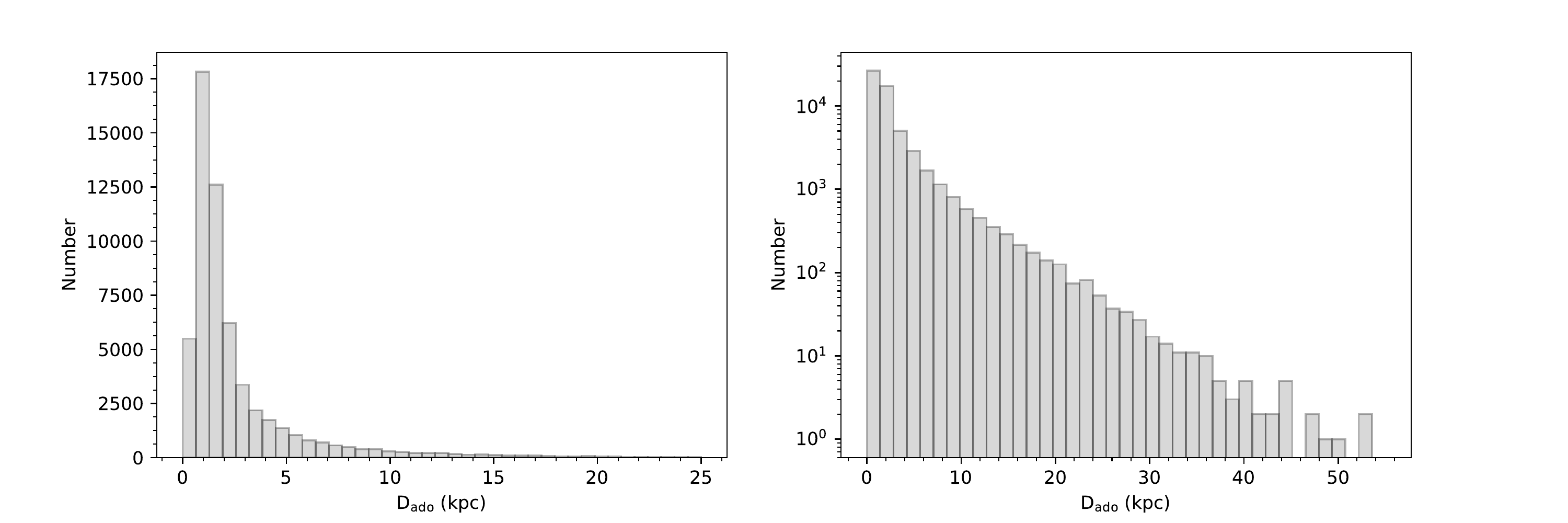}
\caption{Distributions of the adopted (photometric) distance, D$_{\rm ado}$.  The left-hand panel shows a linear vertical scale, while the right-hand panel uses a log vertical scale.  \label{fdistance}}
\end{figure*}

\section{Derivation of Distance Estimates}

Distances to individual stars are estimated using the distance modulus relationships between M$_{v}$ and $(B-V)_0$ described by \citet{beers2000}. These relationships require the likely evolutionary status assigned by the n-SSPP to be specified, which is based on the derived surface gravity estimate.  See \citet{beers2012} for a complete discussion of this method. Based on previous tests of this approach, we expect the distances assigned as described above to be accurate to on the order of 15-20\%. For production of the cartographic maps we only considered stars with available distance measurements, resulting in a total of 58,015 stars.  Figure~\ref{fdistance} shows the distribution of derived distance estimates for the AEGIS sample, using both a linear and a log vertical scale.  From inspection of the figure, although AEGIS includes stars as distant as $\sim 50$ kpc, the great majority of the stars $\sim 85$\,\% are located within 5 kpc of the Sun; it is a relatively local sample.

The complete set of spectra and derived parameters for the AEGIS sample are being prepared for public release in due course.

\bibliography{bibliography}

\begin{thebibliography}{}
\expandafter\ifx\csname natexlab\endcsname\relax\def\natexlab#1{#1}\fi

\bibitem[{{Abate} {et~al.}(2015){Abate}, {Pols}, {Izzard}, \&
  {Karakas}}]{abate2015c}
{Abate}, C., {Pols}, O.~R., {Izzard}, R.~G., \& {Karakas}, A.~I. 2015, \aap,
  581, A22

\bibitem[{{Abbott} {et~al.}(2017){Abbott}, {Abbott}, {Abbott}, {Acernese},
  {Ackley}, {Adams}, {Adams}, {Addesso}, {Adhikari}, {Adya}, \&
  et~al.}]{abbott2017a}
{Abbott}, B.~P., {Abbott}, R., {Abbott}, T.~D., {et~al.} 2017, \apjl, 848, L13

\bibitem[{{Allende Prieto} {et~al.}(2008){Allende Prieto}, {Sivarani}, {Beers},
  {Lee}, {Koesterke}, {Shetrone}, {Sneden}, {Lambert}, {Wilhelm}, {Rockosi},
  {Lai}, {Yanny}, {Ivans}, {Johnson}, {Aoki}, {Bailer-Jones}, \& {Re
  Fiorentin}}]{allende2008}
{Allende Prieto}, C., {Sivarani}, T., {Beers}, T.~C., {et~al.} 2008, \aj, 136,
  2070

\bibitem[{{Amorisco}(2017)}]{amorisco2017}
{Amorisco}, N.~C. 2017, \mnras, 464, 2882

\bibitem[{{An} {et~al.}(2015){An}, {Beers}, {Santucci}, {Carollo}, {Placco},
  {Lee}, \& {Rossi}}]{an2015}
{An}, D., {Beers}, T.~C., {Santucci}, R.~M., {et~al.} 2015, \apjl, 813, L28

\bibitem[{{An} {et~al.}(2013){An}, {Beers}, {Johnson}, {Pinsonneault}, {Lee},
  {Bovy}, {Ivezi{\'c}}, {Carollo}, \& {Newby}}]{an2013}
{An}, D., {Beers}, T.~C., {Johnson}, J.~A., {et~al.} 2013, \apj, 763, 65

\bibitem[{{Aoki} {et~al.}(2007){Aoki}, {Beers}, {Christlieb}, {Norris}, {Ryan},
  \& {Tsangarides}}]{aoki2007}
{Aoki}, W., {Beers}, T.~C., {Christlieb}, N., {et~al.} 2007, \apj, 655, 492

\bibitem[{{Aoki} {et~al.}(2013){Aoki}, {Beers}, {Lee}, {Honda}, {Ito},
  {Takada-Hidai}, {Frebel}, {Suda}, {Fujimoto}, {Carollo}, \&
  {Sivarani}}]{aoki2013}
{Aoki}, W., {Beers}, T.~C., {Lee}, Y.~S., {et~al.} 2013, \aj, 145, 13

\bibitem[{{Arcones} \& {Thielemann}(2013)}]{arcones2013}
{Arcones}, A., \& {Thielemann}, F.-K. 2013, Journal of Physics G Nuclear
  Physics, 40, 013201

\bibitem[{{Astropy Collaboration} {et~al.}(2013){Astropy Collaboration},
  {Robitaille}, {Tollerud}, {Greenfield}, {Droettboom}, {Bray}, {Aldcroft},
  {Davis}, {Ginsburg}, {Price-Whelan}, {Kerzendorf}, {Conley}, {Crighton},
  {Barbary}, {Muna}, {Ferguson}, {Grollier}, {Parikh}, {Nair}, {Unther},
  {Deil}, {Woillez}, {Conseil}, {Kramer}, {Turner}, {Singer}, {Fox}, {Weaver},
  {Zabalza}, {Edwards}, {Azalee Bostroem}, {Burke}, {Casey}, {Crawford},
  {Dencheva}, {Ely}, {Jenness}, {Labrie}, {Lim}, {Pierfederici}, {Pontzen},
  {Ptak}, {Refsdal}, {Servillat}, \& {Streicher}}]{astropy2013}
{Astropy Collaboration}, {Robitaille}, T.~P., {Tollerud}, E.~J., {et~al.} 2013,
  \aap, 558, A33

\bibitem[{{Barklem} {et~al.}(2005){Barklem}, {Christlieb}, {Beers}, {Hill},
  {Bessell}, {Holmberg}, {Marsteller}, {Rossi}, {Zickgraf}, \&
  {Reimers}}]{barklem2005}
{Barklem}, P.~S., {Christlieb}, N., {Beers}, T.~C., {et~al.} 2005, \aap, 439,
  129

\bibitem[{{Beers} {et~al.}(2000){Beers}, {Chiba}, {Yoshii}, {Platais},
  {Hanson}, {Fuchs}, \& {Rossi}}]{beers2000}
{Beers}, T.~C., {Chiba}, M., {Yoshii}, Y., {et~al.} 2000, \aj, 119, 2866

\bibitem[{{Beers} \& {Christlieb}(2005)}]{beers2005}
{Beers}, T.~C., \& {Christlieb}, N. 2005, \araa, 43, 531

\bibitem[{{Beers} {et~al.}(2014){Beers}, {Norris}, {Placco}, {Lee}, {Rossi},
  {Carollo}, \& {Masseron}}]{beers2014}
{Beers}, T.~C., {Norris}, J.~E., {Placco}, V.~M., {et~al.} 2014, \apj, 794, 58

\bibitem[{{Beers} {et~al.}(1985){Beers}, {Preston}, \& {Shectman}}]{beers1985}
{Beers}, T.~C., {Preston}, G.~W., \& {Shectman}, S.~A. 1985, \aj, 90, 2089

\bibitem[{{Beers} {et~al.}(1992){Beers}, {Preston}, \& {Shectman}}]{beers1992}
---. 1992, \aj, 103, 1987

\bibitem[{{Beers} {et~al.}(2012){Beers}, {Carollo}, {Ivezi{\'c}}, {An},
  {Chiba}, {Norris}, {Freeman}, {Lee}, {Munn}, {Re Fiorentin}, {Sivarani},
  {Wilhelm}, {Yanny}, \& {York}}]{beers2012}
{Beers}, T.~C., {Carollo}, D., {Ivezi{\'c}}, {\v Z}., {et~al.} 2012, \apj, 746,
  34

\bibitem[{{Beers} {et~al.}(2017){Beers}, {Placco}, {Carollo}, {Rossi}, {Lee},
  {Frebel}, {Norris}, {Dietz}, \& {Masseron}}]{beers2017}
{Beers}, T.~C., {Placco}, V.~M., {Carollo}, D., {et~al.} 2017, \apj, 835, 81

\bibitem[{{Bekki} \& {Chiba}(2001)}]{bekki2001}
{Bekki}, K., \& {Chiba}, M. 2001, \apj, 558, 666

\bibitem[{{Bland-Hawthorn} \& {Gerhard}(2016)}]{bland-hawthorn2016}
{Bland-Hawthorn}, J., \& {Gerhard}, O. 2016, \araa, 54, 529

\bibitem[{{Bonifacio} {et~al.}(2015){Bonifacio}, {Caffau}, {Spite}, {Limongi},
  {Chieffi}, {Klessen}, {Fran{\c c}ois}, {Molaro}, {Ludwig}, {Zaggia}, {Spite},
  {Plez}, {Cayrel}, {Christlieb}, {Clark}, {Glover}, {Hammer}, {Koch},
  {Monaco}, {Sbordone}, \& {Steffen}}]{bonifacio2015}
{Bonifacio}, P., {Caffau}, E., {Spite}, M., {et~al.} 2015, \aap, 579, A28

\bibitem[{{Bovy}(2015)}]{bovy2015}
{Bovy}, J. 2015, \apjs, 216, 29

\bibitem[{{Bromm} \& {Loeb}(2003)}]{bromm2003}
{Bromm}, V., \& {Loeb}, A. 2003, \nat, 425, 812

\bibitem[{{Brook} {et~al.}(2003){Brook}, {Kawata}, {Gibson}, \&
  {Flynn}}]{brook2003}
{Brook}, C.~B., {Kawata}, D., {Gibson}, B.~K., \& {Flynn}, C. 2003, \apjl, 585,
  L125

\bibitem[{{Brown} {et~al.}(2001){Brown}, {Cai}, \& {DasGupta}}]{brown2001}
{Brown}, L.~D., {Cai}, T.~T., \& {DasGupta}, A. 2001, Statistical Science, 16,
  101

\bibitem[{{Bullock} \& {Johnston}(2005)}]{bullock2005}
{Bullock}, J.~S., \& {Johnston}, K.~V. 2005, \apj, 635, 931

\bibitem[{{Cameron}(2003)}]{cameron2003}
{Cameron}, A.~G.~W. 2003, \apj, 587, 327

\bibitem[{{Carollo} {et~al.}(2014){Carollo}, {Freeman}, {Beers}, {Placco},
  {Tumlinson}, \& {Martell}}]{carollo2014}
{Carollo}, D., {Freeman}, K., {Beers}, T.~C., {et~al.} 2014, \apj, 788, 180

\bibitem[{{Carollo} {et~al.}(2007){Carollo}, {Beers}, {Lee}, {Chiba}, {Norris},
  {Wilhelm}, {Sivarani}, {Marsteller}, {Munn}, {Bailer-Jones}, {Fiorentin}, \&
  {York}}]{carollo2007}
{Carollo}, D., {Beers}, T.~C., {Lee}, Y.~S., {et~al.} 2007, \nat, 450, 1020

\bibitem[{{Carollo} {et~al.}(2010){Carollo}, {Beers}, {Chiba}, {Norris},
  {Freeman}, {Lee}, {Ivezi{\'c}}, {Rockosi}, \& {Yanny}}]{carollo2010}
{Carollo}, D., {Beers}, T.~C., {Chiba}, M., {et~al.} 2010, \apj, 712, 692

\bibitem[{{Carollo} {et~al.}(2012){Carollo}, {Beers}, {Bovy}, {Sivarani},
  {Norris}, {Freeman}, {Aoki}, {Lee}, \& {Kennedy}}]{carollo2012}
{Carollo}, D., {Beers}, T.~C., {Bovy}, J., {et~al.} 2012, \apj, 744, 195

\bibitem[{{Chiaki} {et~al.}(2017){Chiaki}, {Tominaga}, \&
  {Nozawa}}]{chiaki2017}
{Chiaki}, G., {Tominaga}, N., \& {Nozawa}, T. 2017, \mnras, 472, L115

\bibitem[{{Chiappini}(2013)}]{chiappini2013}
{Chiappini}, C. 2013, Astronomische Nachrichten, 334, 595

\bibitem[{{Chiba} \& {Beers}(2000)}]{chiba2000}
{Chiba}, M., \& {Beers}, T.~C. 2000, \aj, 119, 2843

\bibitem[{{Choplin} {et~al.}(2017){Choplin}, {Hirschi}, {Meynet}, \&
  {Ekstr{\"o}m}}]{choplin2017}
{Choplin}, A., {Hirschi}, R., {Meynet}, G., \& {Ekstr{\"o}m}, S. 2017, \aap,
  607, L3

\bibitem[{{Christlieb}(2003)}]{christlieb2003}
{Christlieb}, N. 2003, in Reviews in Modern Astronomy, Vol.~16, Reviews in
  Modern Astronomy, ed. R.~E. {Schielicke}, 191

\bibitem[{{Christlieb} {et~al.}(2004){Christlieb}, {Beers}, {Barklem},
  {Bessell}, {Hill}, {Holmberg}, {Korn}, {Marsteller}, {Mashonkina}, {Qian},
  {Rossi}, {Wasserburg}, {Zickgraf}, {Kratz}, {Nordstr{\"o}m}, {Pfeiffer},
  {Rhee}, \& {Ryan}}]{christlieb2004}
{Christlieb}, N., {Beers}, T.~C., {Barklem}, P.~S., {et~al.} 2004, \aap, 428,
  1027

\bibitem[{{Cohen} {et~al.}(2005){Cohen}, {Shectman}, {Thompson}, {McWilliam},
  {Christlieb}, {Melendez}, {Zickgraf}, {Ram{\'{\i}}rez}, \&
  {Swenson}}]{cohen2005}
{Cohen}, J.~G., {Shectman}, S., {Thompson}, I., {et~al.} 2005, \apjl, 633, L109

\bibitem[{{C{\^o}t{\'e}} {et~al.}(2016){C{\^o}t{\'e}}, {Ritter}, {O'Shea},
  {Herwig}, {Pignatari}, {Jones}, \& {Fryer}}]{cote2016}
{C{\^o}t{\'e}}, B., {Ritter}, C., {O'Shea}, B.~W., {et~al.} 2016, \apj, 824, 82

\bibitem[{{Cowan} \& {Rose}(1977)}]{cowan1977}
{Cowan}, J.~J., \& {Rose}, W.~K. 1977, \apj, 212, 149

\bibitem[{{Cui} {et~al.}(2012){Cui}, {Zhao}, {Chu}, {Li}, {Li}, {Zhang}, {Su},
  {Yao}, {Wang}, {Xing}, {Li}, {Zhu}, {Wang}, {Gu}, {Luo}, {Xu}, {Zhang},
  {Liu}, {Zhang}, {Yang}, {Cao}, {Chen}, {Chen}, {Chen}, {Chen}, {Chu}, {Feng},
  {Gong}, {Hou}, {Hu}, {Hu}, {Hu}, {Jia}, {Jiang}, {Jiang}, {Jiang}, {Jin},
  {Li}, {Li}, {Li}, {Liu}, {Liu}, {Lu}, {Mao}, {Men}, {Qi}, {Qi}, {Shi},
  {Tang}, {Tao}, {Wang}, {Wang}, {Wang}, {Wang}, {Wang}, {Wang}, {Wang},
  {Wang}, {Wang}, {Wang}, {Wang}, {Wang}, {Xu}, {Xu}, {Yang}, {Yu}, {Yuan},
  {Yuan}, {Zhai}, {Zhang}, {Zhang}, {Zhang}, {Zhao}, {Zhou}, {Zhou}, {Zhu}, \&
  {Zou}}]{cui2012}
{Cui}, X.-Q., {Zhao}, Y.-H., {Chu}, Y.-Q., {et~al.} 2012, Research in Astronomy
  and Astrophysics, 12, 1197

\bibitem[{{Dardelet} {et~al.}(2015){Dardelet}, {Ritter}, {Prado}, {Heringer},
  {Higgs}, {Sandalski}, {Jones}, {Denissenkov}, {Venn}, {Bertolli},
  {Pignatari}, {Woodward}, \& {Herwig}}]{dardelet2015}
{Dardelet}, L., {Ritter}, C., {Prado}, P., {et~al.} 2015, ArXiv e-prints,
  arXiv:1505.05500

\bibitem[{{Das} \& {Binney}(2016)}]{das2016}
{Das}, P., \& {Binney}, J. 2016, \mnras, 460, 1725

\bibitem[{{de Bennassuti} {et~al.}(2017){de Bennassuti}, {Salvadori},
  {Schneider}, {Valiante}, \& {Omukai}}]{debennassuti2017}
{de Bennassuti}, M., {Salvadori}, S., {Schneider}, R., {Valiante}, R., \&
  {Omukai}, K. 2017, \mnras, 465, 926

\bibitem[{{de Jong} {et~al.}(2010){de Jong}, {Yanny}, {Rix}, {Dolphin},
  {Martin}, \& {Beers}}]{dejong2010}
{de Jong}, J.~T.~A., {Yanny}, B., {Rix}, H.-W., {et~al.} 2010, \apj, 714, 663

\bibitem[{{De Lucia} \& {Helmi}(2008)}]{delucia2008}
{De Lucia}, G., \& {Helmi}, A. 2008, \mnras, 391, 14

\bibitem[{{Diemand} {et~al.}(2005){Diemand}, {Madau}, \& {Moore}}]{diemand2005}
{Diemand}, J., {Madau}, P., \& {Moore}, B. 2005, \mnras, 364, 367

\bibitem[{{Drout} {et~al.}(2017){Drout}, {Piro}, {Shappee}, {Kilpatrick},
  {Simon}, {Contreras}, {Coulter}, {Foley}, {Siebert}, {Morrell}, {Boutsia},
  {Di Mille}, {Holoien}, {Kasen}, {Kollmeier}, {Madore}, {Monson},
  {Murguia-Berthier}, {Pan}, {Prochaska}, {Ramirez-Ruiz}, {Rest}, {Adams},
  {Alatalo}, {Ba{\~n}ados}, {Baughman}, {Beers}, {Bernstein}, {Bitsakis},
  {Campillay}, {Hansen}, {Higgs}, {Ji}, {Maravelias}, {Marshall}, {Bidin},
  {Prieto}, {Rasmussen}, {Rojas-Bravo}, {Strom}, {Ulloa},
  {Vargas-Gonz{\'a}lez}, {Wan}, \& {Whitten}}]{drout2017}
{Drout}, M.~R., {Piro}, A.~L., {Shappee}, B.~J., {et~al.} 2017, Science, 358,
  1570

\bibitem[{{Eggen} {et~al.}(1962){Eggen}, {Lynden-Bell}, \&
  {Sandage}}]{eggen1962}
{Eggen}, O.~J., {Lynden-Bell}, D., \& {Sandage}, A.~R. 1962, \apj, 136, 748

\bibitem[{{El-Badry} {et~al.}(2018){El-Badry}, {Bland-Hawthorn}, {Wetzel},
  {Quataert}, {Weisz}, {Boylan-Kolchin}, {Hopkins}, {Faucher-Gigu{\`e}re},
  {Kere{\v s}}, \& {Garrison-Kimmel}}]{el-badry2018}
{El-Badry}, K., {Bland-Hawthorn}, J., {Wetzel}, A., {et~al.} 2018, ArXiv
  e-prints, arXiv:1804.00659

\bibitem[{{Frebel}(2018)}]{frebel2018}
{Frebel}, A. 2018, Annu. Rev. Nucl. Part. S., in press

\bibitem[{{Frebel} {et~al.}(2013){Frebel}, {Casey}, {Jacobson}, \&
  {Yu}}]{frebel2013}
{Frebel}, A., {Casey}, A.~R., {Jacobson}, H.~R., \& {Yu}, Q. 2013, \apj, 769,
  57

\bibitem[{{Frebel} {et~al.}(2006){Frebel}, {Christlieb}, {Norris}, {Aoki}, \&
  {Asplund}}]{frebel2006b}
{Frebel}, A., {Christlieb}, N., {Norris}, J.~E., {Aoki}, W., \& {Asplund}, M.
  2006, \apjl, 638, L17

\bibitem[{{Frebel} {et~al.}(2008){Frebel}, {Collet}, {Eriksson}, {Christlieb},
  \& {Aoki}}]{frebel2008}
{Frebel}, A., {Collet}, R., {Eriksson}, K., {Christlieb}, N., \& {Aoki}, W.
  2008, \apj, 684, 588

\bibitem[{{Frebel} {et~al.}(2007){Frebel}, {Johnson}, \& {Bromm}}]{frebel2007b}
{Frebel}, A., {Johnson}, J.~L., \& {Bromm}, V. 2007, \mnras, 380, L40

\bibitem[{{Frischknecht} {et~al.}(2016){Frischknecht}, {Hirschi}, {Pignatari},
  {Maeder}, {Meynet}, {Chiappini}, {Thielemann}, {Rauscher}, {Georgy}, \&
  {Ekstr{\"o}m}}]{frischknecht2016}
{Frischknecht}, U., {Hirschi}, R., {Pignatari}, M., {et~al.} 2016, \mnras, 456,
  1803

\bibitem[{{Frost} \& {Lattanzio}(1996)}]{frost1996}
{Frost}, C., \& {Lattanzio}, J. 1996, ArXiv Astrophysics e-prints,
  astro-ph/9601017

\bibitem[{{Fujimoto} {et~al.}(2008){Fujimoto}, {Nishimura}, \&
  {Hashimoto}}]{fujimoto2008}
{Fujimoto}, S.-i., {Nishimura}, N., \& {Hashimoto}, M.-a. 2008, \apj, 680, 1350

\bibitem[{{Hampel} {et~al.}(2016){Hampel}, {Stancliffe}, {Lugaro}, \&
  {Meyer}}]{hampel2016}
{Hampel}, M., {Stancliffe}, R.~J., {Lugaro}, M., \& {Meyer}, B.~S. 2016, \apj,
  831, 171

\bibitem[{{Hansen} {et~al.}(2016{\natexlab{a}}){Hansen}, {Nordstr{\"o}m},
  {Hansen}, {Kennedy}, {Placco}, {Beers}, {Andersen}, {Cescutti}, \&
  {Chiappini}}]{chansen2016}
{Hansen}, C.~J., {Nordstr{\"o}m}, B., {Hansen}, T.~T., {et~al.}
  2016{\natexlab{a}}, \aap, 588, A37

\bibitem[{{Hansen} {et~al.}(2016{\natexlab{b}}){Hansen}, {Andersen},
  {Nordstr{\"o}m}, {Beers}, {Placco}, {Yoon}, \& {Buchhave}}]{hansen2016a}
{Hansen}, T.~T., {Andersen}, J., {Nordstr{\"o}m}, B., {et~al.}
  2016{\natexlab{b}}, \aap, 586, A160

\bibitem[{{Hansen} {et~al.}(2016{\natexlab{c}}){Hansen}, {Andersen},
  {Nordstr{\"o}m}, {Beers}, {Placco}, {Yoon}, \& {Buchhave}}]{hansen2016b}
---. 2016{\natexlab{c}}, \aap, 588, A3

\bibitem[{{Hartwig} {et~al.}(2018){Hartwig}, {Yoshida}, {Magg}, {Frebel},
  {Glover}, {G{\'o}mez}, {Griffen}, {Ishigaki}, {Ji}, {Klessen}, {O'Shea}, \&
  {Tominaga}}]{hartwig2018}
{Hartwig}, T., {Yoshida}, N., {Magg}, M., {et~al.} 2018, ArXiv e-prints,
  arXiv:1801.05044

\bibitem[{{Henden} {et~al.}(2015){Henden}, {Levine}, {Terrell}, \&
  {Welch}}]{henden2015}
{Henden}, A.~A., {Levine}, S., {Terrell}, D., \& {Welch}, D.~L. 2015, in
  American Astronomical Society Meeting Abstracts, Vol. 225, American
  Astronomical Society Meeting Abstracts \#225, 336.16

\bibitem[{{Herwig}(2005)}]{herwig2005}
{Herwig}, F. 2005, \araa, 43, 435

\bibitem[{{Honda} {et~al.}(2007){Honda}, {Aoki}, {Ishimaru}, \&
  {Wanajo}}]{honda2007}
{Honda}, S., {Aoki}, W., {Ishimaru}, Y., \& {Wanajo}, S. 2007, \apj, 666, 1189

\bibitem[{Hunter(2007)}]{hunter2007}
Hunter, J.~D. 2007, Computing In Science \& Engineering, 9, 90

\bibitem[{{Ishigaki} {et~al.}(2018){Ishigaki}, {Tominaga}, {Kobayashi}, \&
  {Nomoto}}]{ishigaki2018}
{Ishigaki}, M.~N., {Tominaga}, N., {Kobayashi}, C., \& {Nomoto}, K. 2018, \apj,
  857, 46

\bibitem[{{Izutani} {et~al.}(2009){Izutani}, {Umeda}, \&
  {Tominaga}}]{izutani2009}
{Izutani}, N., {Umeda}, H., \& {Tominaga}, N. 2009, \apj, 692, 1517

\bibitem[{{Jacobson} {et~al.}(2015){Jacobson}, {Keller}, {Frebel}, {Casey},
  {Asplund}, {Bessell}, {Da Costa}, {Lind}, {Marino}, {Norris}, {Pe{\~n}a},
  {Schmidt}, {Tisserand}, {Walsh}, {Yong}, \& {Yu}}]{jacobson2015}
{Jacobson}, H.~R., {Keller}, S., {Frebel}, A., {et~al.} 2015, \apj, 807, 171

\bibitem[{{Jeon} {et~al.}(2017){Jeon}, {Besla}, \& {Bromm}}]{jeon2017}
{Jeon}, M., {Besla}, G., \& {Bromm}, V. 2017, \apj, 848, 85

\bibitem[{Jones {et~al.}(2001)Jones, Oliphant, Peterson, {et~al.}}]{scipy2001}
Jones, E., Oliphant, T., Peterson, P., {et~al.} 2001, {SciPy}: Open source
  scientific tools for {Python}, ,

\bibitem[{{Karakas} \& {Lattanzio}(2014)}]{karakas2014}
{Karakas}, A.~I., \& {Lattanzio}, J.~C. 2014, \pasa, 31, e030

\bibitem[{{Keller} {et~al.}(2007){Keller}, {Schmidt}, {Bessell}, {Conroy},
  {Francis}, {Granlund}, {Kowald}, {Oates}, {Martin-Jones}, {Preston},
  {Tisserand}, \& {Vaccarella}}]{keller2007}
{Keller}, S.~C., {Schmidt}, B.~P., {Bessell}, M.~S., {et~al.} 2007, \pasa, 24,
  1

\bibitem[{{Keller} {et~al.}(2014){Keller}, {Bessell}, {Frebel}, {Casey},
  {Asplund}, {Jacobson}, {Lind}, {Norris}, {Yong}, {Heger}, {Magic}, {da
  Costa}, {Schmidt}, \& {Tisserand}}]{keller2014}
{Keller}, S.~C., {Bessell}, M.~S., {Frebel}, A., {et~al.} 2014, \nat, 506, 463

\bibitem[{{Kobayashi} {et~al.}(2011){Kobayashi}, {Tominaga}, \&
  {Nomoto}}]{kobayashi2011}
{Kobayashi}, C., {Tominaga}, N., \& {Nomoto}, K. 2011, \apjl, 730, L14

\bibitem[{{Lattimer} \& {Schramm}(1974)}]{lattimer1974}
{Lattimer}, J.~M., \& {Schramm}, D.~N. 1974, \apjl, 192, L145

\bibitem[{{Lee} {et~al.}(2017){Lee}, {Beers}, {Kim}, {Placco}, {Yoon},
  {Carollo}, {Masseron}, \& {Jung}}]{lee2017}
{Lee}, Y.~S., {Beers}, T.~C., {Kim}, Y.~K., {et~al.} 2017, \apj, 836, 91

\bibitem[{{Lee} {et~al.}(2014){Lee}, {Suda}, {Beers}, \&
  {Stancliffe}}]{lee2014}
{Lee}, Y.~S., {Suda}, T., {Beers}, T.~C., \& {Stancliffe}, R.~J. 2014, \apj,
  788, 131

\bibitem[{{Lee} {et~al.}(2008{\natexlab{a}}){Lee}, {Beers}, {Sivarani},
  {Allende Prieto}, {Koesterke}, {Wilhelm}, {Re Fiorentin}, {Bailer-Jones},
  {Norris}, {Rockosi}, {Yanny}, {Newberg}, {Covey}, {Zhang}, \&
  {Luo}}]{lee2008a}
{Lee}, Y.~S., {Beers}, T.~C., {Sivarani}, T., {et~al.} 2008{\natexlab{a}}, \aj,
  136, 2022

\bibitem[{{Lee} {et~al.}(2008{\natexlab{b}}){Lee}, {Beers}, {Sivarani},
  {Johnson}, {An}, {Wilhelm}, {Allende Prieto}, {Koesterke}, {Re Fiorentin},
  {Bailer-Jones}, {Norris}, {Yanny}, {Rockosi}, {Newberg}, {Cudworth}, \&
  {Pan}}]{lee2008b}
---. 2008{\natexlab{b}}, \aj, 136, 2050

\bibitem[{{Lee} {et~al.}(2011){Lee}, {Beers}, {Allende Prieto}, {Lai},
  {Rockosi}, {Morrison}, {Johnson}, {An}, {Sivarani}, \& {Yanny}}]{lee2011}
{Lee}, Y.~S., {Beers}, T.~C., {Allende Prieto}, C., {et~al.} 2011, \aj, 141, 90

\bibitem[{{Lee} {et~al.}(2013){Lee}, {Beers}, {Masseron}, {Plez}, {Rockosi},
  {Sobeck}, {Yanny}, {Lucatello}, {Sivarani}, {Placco}, \& {Carollo}}]{lee2013}
{Lee}, Y.~S., {Beers}, T.~C., {Masseron}, T., {et~al.} 2013, \aj, 146, 132

\bibitem[{{Lucatello} {et~al.}(2006){Lucatello}, {Beers}, {Christlieb},
  {Barklem}, {Rossi}, {Marsteller}, {Sivarani}, \& {Lee}}]{lucatello2006}
{Lucatello}, S., {Beers}, T.~C., {Christlieb}, N., {et~al.} 2006, \apjl, 652,
  L37

\bibitem[{{Lugaro} {et~al.}(2003){Lugaro}, {Herwig}, {Lattanzio}, {Gallino}, \&
  {Straniero}}]{lugaro2003}
{Lugaro}, M., {Herwig}, F., {Lattanzio}, J.~C., {Gallino}, R., \& {Straniero},
  O. 2003, \apj, 586, 1305

\bibitem[{{Lugaro} {et~al.}(2012){Lugaro}, {Karakas}, {Stancliffe}, \&
  {Rijs}}]{lugaro2012}
{Lugaro}, M., {Karakas}, A.~I., {Stancliffe}, R.~J., \& {Rijs}, C. 2012, \apj,
  747, 2

\bibitem[{{Maeder} \& {Meynet}(2015)}]{maeder2015}
{Maeder}, A., \& {Meynet}, G. 2015, \aap, 580, A32

\bibitem[{{McWilliam}(1997)}]{mcwilliam1997}
{McWilliam}, A. 1997, \araa, 35, 503

\bibitem[{{Meyer}(1989)}]{meyer1989}
{Meyer}, B.~S. 1989, \apj, 343, 254

\bibitem[{{Meynet} {et~al.}(2006){Meynet}, {Ekstr{\"o}m}, \&
  {Maeder}}]{meynet2006}
{Meynet}, G., {Ekstr{\"o}m}, S., \& {Maeder}, A. 2006, \aap, 447, 623

\bibitem[{{Meynet} {et~al.}(2010){Meynet}, {Hirschi}, {Ekstrom}, {Maeder},
  {Georgy}, {Eggenberger}, \& {Chiappini}}]{meynet2010}
{Meynet}, G., {Hirschi}, R., {Ekstrom}, S., {et~al.} 2010, \aap, 521, A30

\bibitem[{{Nomoto} {et~al.}(2013){Nomoto}, {Kobayashi}, \&
  {Tominaga}}]{nomoto2013}
{Nomoto}, K., {Kobayashi}, C., \& {Tominaga}, N. 2013, \araa, 51, 457

\bibitem[{{Norris} {et~al.}(2013){Norris}, {Yong}, {Bessell}, {Christlieb},
  {Asplund}, {Gilmore}, {Wyse}, {Beers}, {Barklem}, {Frebel}, \&
  {Ryan}}]{norris2013b}
{Norris}, J.~E., {Yong}, D., {Bessell}, M.~S., {et~al.} 2013, \apj, 762, 28

\bibitem[{{Omukai} {et~al.}(2005){Omukai}, {Tsuribe}, {Schneider}, \&
  {Ferrara}}]{omukai2005}
{Omukai}, K., {Tsuribe}, T., {Schneider}, R., \& {Ferrara}, A. 2005, \apj, 626,
  627

\bibitem[{{Placco} {et~al.}(2016{\natexlab{a}}){Placco}, {Beers}, {Reggiani},
  \& {Mel{\'e}ndez}}]{placco2016a}
{Placco}, V.~M., {Beers}, T.~C., {Reggiani}, H., \& {Mel{\'e}ndez}, J.
  2016{\natexlab{a}}, \apjl, 829, L24

\bibitem[{{Placco} {et~al.}(2013){Placco}, {Frebel}, {Beers}, {Karakas},
  {Kennedy}, {Rossi}, {Christlieb}, \& {Stancliffe}}]{placco2013}
{Placco}, V.~M., {Frebel}, A., {Beers}, T.~C., {et~al.} 2013, \apj, 770, 104

\bibitem[{{Placco} {et~al.}(2014){Placco}, {Frebel}, {Beers}, \&
  {Stancliffe}}]{placco2014c}
{Placco}, V.~M., {Frebel}, A., {Beers}, T.~C., \& {Stancliffe}, R.~J. 2014,
  \apj, 797, 21

\bibitem[{{Placco} {et~al.}(2015){Placco}, {Frebel}, {Lee}, {Jacobson},
  {Beers}, {Pena}, {Chan}, \& {Heger}}]{placco2015}
{Placco}, V.~M., {Frebel}, A., {Lee}, Y.~S., {et~al.} 2015, \apj, 809, 136

\bibitem[{{Placco} {et~al.}(2016{\natexlab{b}}){Placco}, {Frebel}, {Beers},
  {Yoon}, {Chiti}, {Heger}, {Chan}, {Casey}, \& {Christlieb}}]{placco2016b}
{Placco}, V.~M., {Frebel}, A., {Beers}, T.~C., {et~al.} 2016{\natexlab{b}},
  \apj, 833, 21

\bibitem[{{Roederer} {et~al.}(2014){Roederer}, {Preston}, {Thompson},
  {Shectman}, {Sneden}, {Burley}, \& {Kelson}}]{roederer2014}
{Roederer}, I.~U., {Preston}, G.~W., {Thompson}, I.~B., {et~al.} 2014, \aj,
  147, 136

\bibitem[{{Rosswog} {et~al.}(2014){Rosswog}, {Korobkin}, {Arcones},
  {Thielemann}, \& {Piran}}]{rosswog2014}
{Rosswog}, S., {Korobkin}, O., {Arcones}, A., {Thielemann}, F.-K., \& {Piran},
  T. 2014, \mnras, 439, 744

\bibitem[{{Salvadori} {et~al.}(2016){Salvadori}, {Sk{\'u}lad{\'o}ttir}, \& {de
  Bennassuti}}]{salvadori2016}
{Salvadori}, S., {Sk{\'u}lad{\'o}ttir}, {\'A}., \& {de Bennassuti}, M. 2016,
  Astronomische Nachrichten, 337, 935

\bibitem[{{Salvadori} {et~al.}(2015){Salvadori}, {Sk{\'u}lad{\'o}ttir}, \&
  {Tolstoy}}]{salvadori2015}
{Salvadori}, S., {Sk{\'u}lad{\'o}ttir}, {\'A}., \& {Tolstoy}, E. 2015, \mnras,
  454, 1320

\bibitem[{{Samland} \& {Gerhard}(2003)}]{samland2003}
{Samland}, M., \& {Gerhard}, O.~E. 2003, \aap, 399, 961

\bibitem[{{Sarmento} {et~al.}(2017){Sarmento}, {Scannapieco}, \&
  {Pan}}]{sarmento2017}
{Sarmento}, R., {Scannapieco}, E., \& {Pan}, L. 2017, \apj, 834, 23

\bibitem[{{Schlegel} {et~al.}(1998){Schlegel}, {Finkbeiner}, \&
  {Davis}}]{schlegel1998}
{Schlegel}, D.~J., {Finkbeiner}, D.~P., \& {Davis}, M. 1998, \apj, 500, 525

\bibitem[{{Schneider} {et~al.}(2003){Schneider}, {Ferrara}, {Salvaterra},
  {Omukai}, \& {Bromm}}]{schneider2003}
{Schneider}, R., {Ferrara}, A., {Salvaterra}, R., {Omukai}, K., \& {Bromm}, V.
  2003, \nat, 422, 869

\bibitem[{{Schneider} {et~al.}(2012){Schneider}, {Omukai}, {Bianchi}, \&
  {Valiante}}]{schneider2012}
{Schneider}, R., {Omukai}, K., {Bianchi}, S., \& {Valiante}, R. 2012, \mnras,
  419, 1566

\bibitem[{{Searle} \& {Zinn}(1978)}]{searle1978}
{Searle}, L., \& {Zinn}, R. 1978, \apj, 225, 357

\bibitem[{{Shappee} {et~al.}(2017){Shappee}, {Simon}, {Drout}, {Piro},
  {Morrell}, {Prieto}, {Kasen}, {Holoien}, {Kollmeier}, {Kelson}, {Coulter},
  {Foley}, {Kilpatrick}, {Siebert}, {Madore}, {Murguia-Berthier}, {Pan},
  {Prochaska}, {Ramirez-Ruiz}, {Rest}, {Adams}, {Alatalo}, {Ba{\~n}ados},
  {Baughman}, {Bernstein}, {Bitsakis}, {Boutsia}, {Bravo}, {Di Mille}, {Higgs},
  {Ji}, {Maravelias}, {Marshall}, {Placco}, {Prieto}, \& {Wan}}]{shappee2017}
{Shappee}, B.~J., {Simon}, J.~D., {Drout}, M.~R., {et~al.} 2017, Science, 358,
  1574

\bibitem[{{Sharma} {et~al.}(2017){Sharma}, {Theuns}, \& {Frenk}}]{sharma2017}
{Sharma}, M., {Theuns}, T., \& {Frenk}, C. 2017, ArXiv e-prints,
  arXiv:1712.05811

\bibitem[{{Sharp} {et~al.}(2006){Sharp}, {Saunders}, {Smith}, {Churilov},
  {Correll}, {Dawson}, {Farrel}, {Frost}, {Haynes}, {Heald}, {Lankshear},
  {Mayfield}, {Waller}, \& {Whittard}}]{sharp2006}
{Sharp}, R., {Saunders}, W., {Smith}, G., {et~al.} 2006, in \procspie, Vol.
  6269, Society of Photo-Optical Instrumentation Engineers (SPIE) Conference
  Series, 62690G

\bibitem[{{Skrutskie} {et~al.}(2006){Skrutskie}, {Cutri}, {Stiening},
  {Weinberg}, {Schneider}, {Carpenter}, {Beichman}, {Capps}, {Chester}, \&
  {Elias}}]{skrutskie2006}
{Skrutskie}, M.~F., {Cutri}, R.~M., {Stiening}, R., {et~al.} 2006, \aj, 131,
  1163

\bibitem[{{Smolinski} {et~al.}(2011){Smolinski}, {Lee}, {Beers}, {An},
  {Bickerton}, {Johnson}, {Loomis}, {Rockosi}, {Sivarani}, \&
  {Yanny}}]{smolinski2011}
{Smolinski}, J.~P., {Lee}, Y.~S., {Beers}, T.~C., {et~al.} 2011, \aj, 141, 89

\bibitem[{{Spite} {et~al.}(2013){Spite}, {Caffau}, {Bonifacio}, {Spite},
  {Ludwig}, {Plez}, \& {Christlieb}}]{spite2013}
{Spite}, M., {Caffau}, E., {Bonifacio}, P., {et~al.} 2013, \aap, 552, A107

\bibitem[{{Starkenburg} {et~al.}(2017){Starkenburg}, {Oman}, {Navarro},
  {Crain}, {Fattahi}, {Frenk}, {Sawala}, \& {Schaye}}]{starkenburg2017}
{Starkenburg}, E., {Oman}, K.~A., {Navarro}, J.~F., {et~al.} 2017, \mnras, 465,
  2212

\bibitem[{{Steinmetz} \& {Muller}(1995)}]{steinmetz1995}
{Steinmetz}, M., \& {Muller}, E. 1995, \mnras, 276, 549

\bibitem[{{Suda} {et~al.}(2011){Suda}, {Yamada}, {Katsuta}, {Komiya},
  {Ishizuka}, {Aoki}, \& {Fujimoto}}]{suda2011}
{Suda}, T., {Yamada}, S., {Katsuta}, Y., {et~al.} 2011, \mnras, 412, 843

\bibitem[{{Suda} {et~al.}(2013){Suda}, {Komiya}, {Yamada}, {Katsuta}, {Aoki},
  {Gil-Pons}, {Doherty}, {Campbell}, {Wood}, \& {Fujimoto}}]{suda2013}
{Suda}, T., {Komiya}, Y., {Yamada}, S., {et~al.} 2013, \mnras, 432, L46

\bibitem[{{Tissera} {et~al.}(2014){Tissera}, {Beers}, {Carollo}, \&
  {Scannapieco}}]{tissera2014}
{Tissera}, P.~B., {Beers}, T.~C., {Carollo}, D., \& {Scannapieco}, C. 2014,
  \mnras, 439, 3128

\bibitem[{{Tominaga} {et~al.}(2014){Tominaga}, {Iwamoto}, \&
  {Nomoto}}]{tominaga2014}
{Tominaga}, N., {Iwamoto}, N., \& {Nomoto}, K. 2014, \apj, 785, 98

\bibitem[{{Travaglio} {et~al.}(2004){Travaglio}, {Gallino}, {Arnone}, {Cowan},
  {Jordan}, \& {Sneden}}]{travaglio2004}
{Travaglio}, C., {Gallino}, R., {Arnone}, E., {et~al.} 2004, \apj, 601, 864

\bibitem[{{Tumlinson}(2007)}]{tumlinson2007b}
{Tumlinson}, J. 2007, \apj, 665, 1361

\bibitem[{{Umeda} \& {Nomoto}(2003)}]{umeda2003}
{Umeda}, H., \& {Nomoto}, K. 2003, \nat, 422, 871

\bibitem[{{Umeda} \& {Nomoto}(2005)}]{umeda2005}
---. 2005, \apj, 619, 427

\bibitem[{{Van Der Walt} {et~al.}(2011){Van Der Walt}, {Colbert}, \&
  {Varoquaux}}]{numpy}
{Van Der Walt}, S., {Colbert}, S.~C., \& {Varoquaux}, G. 2011, ArXiv e-prints,
  arXiv:1102.1523

\bibitem[{{Wanajo} \& {Ishimaru}(2006)}]{wanajo2006}
{Wanajo}, S., \& {Ishimaru}, Y. 2006, Nuclear Physics A, 777, 676

\bibitem[{Wilson(1927)}]{wilson1927}
Wilson, E.~B. 1927, Journal of the American Statistical Association, 22, 209

\bibitem[{{Winteler} {et~al.}(2012){Winteler}, {K{\"a}ppeli}, {Perego},
  {Arcones}, {Vasset}, {Nishimura}, {Liebend{\"o}rfer}, \&
  {Thielemann}}]{winteler2012}
{Winteler}, C., {K{\"a}ppeli}, R., {Perego}, A., {et~al.} 2012, \apjl, 750, L22

\bibitem[{{Wolf} {et~al.}(2018){Wolf}, {Onken}, {Luvaul}, {Schmidt}, {Bessell},
  {Chang}, {Da Costa}, {Mackey}, {Martin-Jones}, {Murphy}, {Preston}, {Scalzo},
  {Shao}, {Smillie}, {Tisserand}, {White}, \& {Yuan}}]{wolf2018}
{Wolf}, C., {Onken}, C.~A., {Luvaul}, L.~C., {et~al.} 2018, \pasa, 35, e010

\bibitem[{{Xu} {et~al.}(2018){Xu}, {Liu}, {Xue}, {Newberg}, {Carlin}, {Xia},
  {Deng}, {Li}, {Zhang}, {Hou}, {Wang}, \& {Cao}}]{xu2018}
{Xu}, Y., {Liu}, C., {Xue}, X.-X., {et~al.} 2018, \mnras, 473, 1244

\bibitem[{{Yamada} {et~al.}(2013){Yamada}, {Suda}, {Komiya}, {Aoki}, \&
  {Fujimoto}}]{yamada2013}
{Yamada}, S., {Suda}, T., {Komiya}, Y., {Aoki}, W., \& {Fujimoto}, M.~Y. 2013,
  \mnras, 436, 1362

\bibitem[{{Yanny} {et~al.}(2009){Yanny}, {Rockosi}, {Newberg}, {Knapp},
  {Adelman-McCarthy}, {Alcorn}, {Allam}, {Allende Prieto}, {An}, {Anderson},
  {Anderson}, \& {Bailer-Jones}}]{yanny2009}
{Yanny}, B., {Rockosi}, C., {Newberg}, H.~J., {et~al.} 2009, \aj, 137, 4377

\bibitem[{{Yoon} {et~al.}(2016){Yoon}, {Beers}, {Placco}, {Rasmussen},
  {Carollo}, {He}, {Hansen}, {Roederer}, \& {Zeanah}}]{yoon2016}
{Yoon}, J., {Beers}, T.~C., {Placco}, V.~M., {et~al.} 2016, \apj, 833, 20

\bibitem[{{York} {et~al.}(2000){York}, {Adelman}, {Anderson}, {Anderson},
  {Annis}, {Bahcall}, {Bakken}, {Barkhouser}, {Bastian}, {Berman}, {Boroski},
  {Bracker}, \& {Briegel}}]{york2000}
{York}, D.~G., {Adelman}, J., {Anderson}, Jr., J.~E., {et~al.} 2000, \aj, 120,
  1579

\bibitem[{{Zhao} \& {Newberg}(2006)}]{zhao2006}
{Zhao}, C., \& {Newberg}, H.~J. 2006, ArXiv Astrophysics e-prints,
  astro-ph/0612034

\bibitem[{{Zolotov} {et~al.}(2009){Zolotov}, {Willman}, {Brooks}, {Governato},
  {Brook}, {Hogg}, {Quinn}, \& {Stinson}}]{zolotov2009}
{Zolotov}, A., {Willman}, B., {Brooks}, A.~M., {et~al.} 2009, \apj, 702, 1058

\end{thebibliography}

\end{document}